\newcommand{\version}{September 17, 2019}
\documentclass[letterpaper,oneside,11pt,pdftex]{article}
\pdfoutput=1
\usepackage[bindingoffset=0.0cm,textheight=22.6cm,hdivide={*,16cm,*}, vdivide={*,22.5cm,*}]{geometry}
\usepackage{amsmath,amsfonts,amssymb}
\usepackage[pdftex]{graphicx}
\usepackage[margin=0.5cm,font=small]{caption}
\usepackage[T1]{fontenc}
\usepackage[utf8]{inputenc}
\usepackage{lmodern}
 \usepackage[numbers,square,comma,sort&compress]{natbib}

\usepackage[pdftex,bookmarksnumbered=true,breaklinks=true]{hyperref}

\makeatletter
\AtBeginDocument{
  \hypersetup{
    pdfkeywords = {\keywords},
    pdftitle = {\@title},
    pdfauthor = {\@author}
  }
}
\makeatother

\makeatletter

 \@addtoreset{equation}{section}
 \makeatother

\renewcommand{\b}{\beta}
\newcommand{\g}{\gamma}
\renewcommand{\d}{\delta}
\newcommand{\e}{\epsilon}
\newcommand{\vare}{\varepsilon}

\renewcommand{\l}{\lambda}
\newcommand{\m}{\mu}
\newcommand{\n}{\nu}

\newcommand{\rh}{\rho}

\newcommand{\ph}{\phi}

\newcommand{\w}{\omega}

\newcommand{\Ph}{\Phi}

\newcommand{\W}{\Omega}

\newcommand{\ml}{\mathfrak{l}}
\newcommand{\mm}{\mathfrak{m}}
\newcommand{\mn}{\mathfrak{n}}

\newcommand{\be}{\begin{equation}}
\newcommand{\ee}{\end{equation}}
\newcommand{\bes}{\begin{subequations}}
\newcommand{\ees}{\end{subequations}}

\newcommand{\pa}{\partial}

\newcommand{\nn}{\nonumber}
\newcommand{\eqnref}[1]{eq. \eqref{#1}}
\newcommand{\na}{\nabla}

\newcommand{\br}{\mathbf{r}}
\newcommand{\bR}{\mathbf{R}}
\newcommand{\bk}{\mathbf{k}}
\newcommand{\bfe}{\mathbf{e}}

\newcommand{\hbk}{\mathbf{\hat k}}
\newcommand{\bu}{\mathbf{u}}
\newcommand{\bU}{\mathbf{U}}
\newcommand{\bx}{\mathbf{x}}

\newcommand{\bfa}{\hat{a}}
\newcommand{\ha}{\bfa}
\newcommand{\bfk}{\mathbf{k}}
\newcommand{\bfki}{\mathbf{k'}}
\newcommand{\bfq}{\mathbf{q}}
\newcommand{\bfv}{\mathbf{v}}
\newcommand{\bfx}{\mathbf{x}}

\newcommand{\intx}{\int d^3\bfx}
\newcommand{\intk}{\int\!\frac{d^3 \bfk}{(2\pi)^3}} 
\newcommand{\intq}{\int\!\frac{d^3 \bfq}{(2\pi)^3}}
\newcommand{\intki}{\int\!\frac{d^3 \bfki}{(2\pi)^3}}

\newcommand{\elA}{D} 
\newcommand{\elC}{C}
\newcommand{\elD}{D}

\newcommand{\kb}{k_{\txt{B}}}
\newcommand{\ct}{c_{_T}}
\newcommand{\cl}{c_{_L}}

\newcommand{\gt}{\gamma_{_{T}}}
\newcommand{\gl}{\gamma_{_{L}}}
\newcommand{\bt}{\beta_{_{T}}}
\newcommand{\bl}{\beta_{_{L}}}
\newcommand{\wl}{\w_{_L}}
\newcommand{\wt}{\w_{_T}}
\newcommand{\Vc}{V_\txt{c}}

\newcommand{\sdfrac}[2]{\mbox{\small$\displaystyle\frac{#1}{#2}$}}

\newcommand{\txt}[1]{\textrm{#1}}

\newcommand{\abs}[1]{|#1|}
\newcommand{\coleq}{\equiv}

\title{\texorpdfstring{\begin{flushright}
        {\small LA-UR-19-25851}
       \end{flushright}\vspace{2em}}{}%
       Dislocation drag from phonon wind in an isotropic crystal at large velocities}

\author{Daniel N. Blaschke, Emil Mottola, and Dean L. Preston}

\date{\version}

\newcommand{\keywords}{dislocations in crystals, drag coefficient, phonon wind}

\graphicspath{{./}{./figures/}}

\begin{document}

\maketitle

\thispagestyle{empty}
\begin{center}
\vspace{-0.3cm}
Los Alamos National Laboratory\\Los Alamos, NM, 87545, USA
\\[0.5cm]
\ttfamily{E-mail: dblaschke@lanl.gov, emil@lanl.gov, dean@lanl.gov}
\end{center}

\vspace{1.5em}

\begin{abstract}
The anharmonic interaction and scattering of phonons by a moving dislocation, the photon wind, imparts a drag force $v\,B(v, T, \rh)$ on the dislocation.
In early studies the drag coefficient $B$ was computed and experimentally determined only for dislocation velocities $v$ much less than transverse sound 
speed, $\ct$. In this paper we derive analytic expressions for the velocity dependence of $B$ up to $\ct$ in terms of the third-order continuum elastic 
constants of an isotropic crystal, in the continuum Debye approximation, valid for dislocation velocities approaching the sound speed. In so doing 
we point out that the most general form of the third order elastic potential for such a crystal and the dislocation-phonon interaction requires two additional 
elastic constants involving asymmetric local rotational strains, which have been neglected previously. We compute the velocity dependence of the transverse 
phonon wind contribution to $B$ in the range 1\%--90\% $\ct$ for Al, Cu, Fe, and Nb in the isotropic Debye approximation.
The drag coefficient for transverse 
phonons scattering from screw dislocations is finite as $v \rightarrow \ct$, whereas $B$ is divergent for transverse phonons scattering from edge dislocations 
in the same limit. This divergence indicates the breakdown of the Debye approximation and sensitivity of the drag coefficient at very high velocities 
to the microscopic crystalline lattice cutoff. We compare our results to experimental results wherever possible and identify ways to validate and 
further improve the theory of dislocation drag at high velocities with realistic phonon dispersion relations, inclusion of lattice cutoff effects, 
MD simulation data, and more accurate experimental measurements.
\end{abstract}

\newpage
\tableofcontents

\section{Introduction and Outline}
\label{sec:intro}

Dislocations are linear defects in the regular ordered structure of a crystalline lattice. Dislocations can move through the crystal that is subjected
to an applied stress field at a speed that is controlled by two mechanisms. At low to intermediate plastic strain rates ($\leq 10^5\, \txt{s}^{-1}$) 
dislocation mobility is limited by their interactions with immobile (forest) dislocations that result in the formation of dislocation-dislocation nodes 
and short junctions. At finite temperature, the combination of the applied stress and local stress fluctuations arising from atomic oscillations result 
in dissociation of the nodes, and the dislocations may more easily glide. The rate-controlling intersection of non-coplanar, attractive mobile and immobile 
dislocations has traditionally been described by Van't Hoff-Arrhenius thermal activation theory, but this approach breaks down at high strain rates.
The theory was recently generalized to strain rates of nearly $10^{12}\, \txt{s}^{-1}$ \cite{Hunter:2015}.
At high strain rates ($\geq 10^5\, \txt{s}^{-1}$),
in addition to dislocation-dislocation interactions, the second mechanism of viscous drag due to interactions with phonons, i.e. the `phonon wind,'
comes into play. The drag force per unit length of dislocation is $v\,B(v, T, \rh)$ where $B$ is the dislocation drag coefficient, $v$ is the dislocation velocity, 
$T$ is the temperature, and $\rh$ is the material density.
It is this second effect of dislocation drag, applicable to the high strain rate regime, that we study in the present paper.

Both the drag coefficient and the mean velocity of dislocations through the lattice increase with increasing applied stress, and the dislocation velocity
may even approach transverse sound speed in the crystal. The accurate evaluation of the drag coefficient at high strain rates is essential for models of 
single-crystal bulk plasticity, polycrystal plasticity, and ductile failure applicable at high stresses and strain rates. In this high-stress regime, where 
mean dislocation speeds vary from a few percent of transverse sound speed, $\ct$, up to nearly $\ct$, the dominant contribution to the dislocation 
drag coefficient is the scattering of phonons by the moving dislocations; in the rest frame of a moving dislocation the phonons moving past the dislocation 
act as a `phonon wind' opposing its glide through the crystal.
Other dissipative effects, which we do not touch upon in this paper as they are subleading 
in the regimes of interest, are the thermoelastic damping and the radiation damping mechanisms, discussed for example in Ref.~\cite{Nadgornyi:1988}.

At low velocities, \emph{i.e.} velocities less than a few percent of transverse sound speed, the drag coefficient due to phonon wind is roughly constant\footnote{
We note that drag is initially dominated by thermal effects at very low dislocation velocities, and that phonon wind becomes important only at velocities of the order of 1\% transverse sound speed and higher.
}, 
but at higher velocities the drag coefficient increases nonlinearly. We note that existing continuum scale models of dislocation drag assume that the 
dislocation velocity is much smaller than $\ct$. To our knowledge there is currently no theoretical framework available in the literature for the accurate 
calculation of the velocity dependence\footnote{
Many authors estimate the velocity dependence of the drag coefficient by means of ``relativistic'' factors $\propto 1/(1-v^2/c^2)^m$ with different exponents $m$ and a limiting (sound) speed $c$ based on purely empirical arguments which lack a first-principles theoretical framework, see \cite{Clifton:1971,Krasnikov:2010,Barton:2011,Luscher:2016,Austin:2018} and references therein.
}
of the dislocation drag coefficient up to $\ct$.
The main focus of this paper is the development of such a framework based on previous work of Alshits and collaborators~\cite{Alshits:1992}, who derived the small velocity limit to lowest order (with ``small'' referring to a few percent $\ct$).
Thus, earlier first-principles continuum models of dislocation drag due to phonon wind are extended from low velocities to nearly $\ct$. The same framework 
also provides the temperature dependence of the drag coefficient up to velocities in the neighborhood of $\ct$, and is flexible enough to incorporate 
more realistic dispersion curves, as well as more accurate experimental or numerical molecular dynamics (MD) data on third-order elastic constants. 
Moreover, since Bose-Einstein statistics is used for the phonon distribution functions, quantum effects are automatically incorporated.

The present calculation of dislocation drag due to phonon scattering assumes a uniformly moving straight (linear) dislocation with a static core.
However, there are additional drag mechanisms arising from the excitation of the internal degrees of freedom of a moving dislocation.
The alternating stress fields associated with the phonons induce oscillations in the shape of the dislocation around linearity and fluctuations around uniform motion.
This is the well-known flutter effect which gives rise to the emission of elastic waves by the dislocations.
It has been argued by Lothe that the elastic wave emission results in a drag on the dislocations proportional to the velocity \cite{Lothe:1960,Lothe:1962}.
In addition to the flutter effect, the lattice periodicity induces fluctuations in the core structure and dislocation velocity that also give rise to the generation of elastic waves and drag on the dislocations.
In order to determine the importance of both effects relative to phonon wind, one needs to study them in the more general quantum field theoretical framework of Alshits et al.~\cite{Alshits:1992}.
This, however, is beyond the scope of the current work where we discuss only phonon wind.

In this paper, which is largely based on the unpublished internal report Ref. \cite{Blaschke:BpaperRpt}, we restrict our study of the phonon wind contribution to the drag coefficient to velocities comparable to but strictly less than $\ct$, 
and dislocation-dislocation interactions are neglected. The case of supersonic dislocations is interesting in its own right, not least because of 
recent MD simulations in fcc metals that indicate the existence of dislocations moving at supersonic speeds; see~\cite{Rosakis:2001,Marian:2006,Pellegrini:2010,Ruestes:2015,Oren:2017} and references therein.
Furthermore, supersonic dislocations have been observed experimentally in plasma crystals~\cite{Nosenko:2007}.
The extension of the theory to include dislocations moving at supersonic speeds, and dislocation-dislocation interactions, will be left for future work.
Crystal anisotropy may have interesting effects on dislocation drag as well; see the recent numerical studies of Refs. \cite{Blaschke:2017lten,Blaschke:2018anis,Blaschke:2019fits}.
However, in focusing on the analytic behavior of dislocation drag up to $\ct$, we presently require the simplifications of the isotropic limit.
In fact, many dislocation-based material strength models for polycrystals \cite{Krasnikov:2010,Barton:2011,Hunter:2015,Austin:2018} make use of the isotropic approximation and as such would benefit from a first-principles derivation of dislocation drag as a function of velocity in the isotropic limit.

Finally, in the course of our derivation we also point out that the most general form of the third-order elastic potential for dislocation-phonon interaction in a crystal in the isotropic limit requires two additional 
elastic constants involving asymmetric local rotational strains, which have been neglected previously based on arguments that only apply to a perfect crystal without dislocations.

The outline of the paper is as follows. In order to provide a self-contained presentation, in Section~\ref{sec:elasticity} we expand the crystal potential in terms 
of displacements from the perfect lattice~\cite{Wallace:1972} to obtain the crystal Hamiltonian.
We then consider a number of approximations and simplifications, such as the restriction to monatomic lattices, and the assumption of material isotropy.
We employ the Debye approximation for the phonon spectrum.
These simplifying approximations enable a semi-analytic approach in which experimentally determined second- and third-order elastic constants are used 
rather than numerical data from classical or quantum MD simulations.
In Section~\ref{sec:dislocations} we discuss the displacement fields of edge and screw dislocations following Eshelby~\cite{Eshelby:1949} and~\cite{Weertman:1980}, and references therein.
Section~\ref{sec:phononwind} is devoted to the phonon wind contribution to the drag coefficient in the continuum approximation.
With the Debye approximation for the phonon spectrum, most of the computation of the drag coefficient can be performed analytically, leaving only a three-dimensional integral that is evaluated numerically.
In Section~\ref{sec:results} we present and discuss our results for a number of metals and compare to experimental values and MD simulations.

\section{Isotropic Solids}
\label{sec:elasticity}
In this section we provide a short review of the elements of continuum elasticity theory pertinent to the calculation of the drag coefficient,
and establish the notation to be used in the subsequent sections.

\subsection{Continuum Hamiltonian and Elastic Constants}
\label{sec:crystalHamiltonian}

The underlying Hamiltonian of a crystalline lattice may be expressed in the form
\begin{equation}
H = \frac{1}{2} \sum_A M^{(A)} \dot \br^{(A)\,2} + \Ph\{\br^{(A)}\}
\label{latHam}
\end{equation}
where $M^{(A)}$ is the mass of the atom at lattice site $A$,  $\{\br^{(A)}\}$ is  the set of all atomic position vectors, 
and $\Ph\{\br^{(A)}\}$ is the crystal potential energy function. Allow the atoms to be displaced 
\begin{equation}
\bu^{(A)} \equiv \br^{(A)} - \bR^{(A)}
\label{udisp}
\end{equation}
with Cartesian components $u_i^{(A)}$ about the set of equilibrium lattice positions $\{\bR^{(A)}\}$,
and expand the potential energy function $\Ph\{\br^{(A)}\}$ in a Taylor series in these displacements.
Taking then the continuum limit, a standard computation leads to the continuum potential energy~\cite{Wallace:1972} 
\begin{equation}
\Ph\{\br\} = \Ph\{\bR\} + \frac{V}{2} \sum_{ijkl} \elC_{ijkl} \,u_{i,j}\,u_{k, l} 
+ \frac{V}{6}\!\sum_{ijklmn} \elD_{ijklmn}\,u_{i,j}\,u_{k,l}\,u_{m,n} + \dots
\label{Phigrad}
\end{equation}
where $\Ph\{\bR\}$ is the potential when all ions are at their equilibrium positions,
and the elastic constants are defined by
\begin{subequations}\label{elasCD}
\begin{align}
\elC_{ijkl} & \equiv \frac{1}{V}\! \sum_{AB} \Ph^{(AB)}_{ik} R^{(A)}_{j} R^{(B)}_{l} \\
\elD_{ijklmn}  & \equiv \frac{1}{V}\!\sum_{ABC} \Ph_{ikm}^{(ABC)} R_{j}^{(A)} R_{l}^{(B)} R_{n}^{(C)}
\end{align}
\end{subequations}
in which 
\begin{subequations}
\begin{align}
\Ph_i^{(A)} &= \frac{\pa \Ph\ } {\pa r_i^{(A)}}\bigg\vert_{\{\br\} = \{\bR\}} = 0 \label{firstPhi}\\
\Ph_{ik}^{(AB)} &= \frac{\pa^2 \Ph\ } {\pa r_i^{(A)}\pa r_k^{(B)}}\bigg\vert_{\{\br\} = \{\bR\}} \\
\Ph_{ikm}^{(ABC)} &= \frac{\pa^3 \Ph\ } {\pa r_i^{(A)}\pa r_k^{(B)}\pa r_m^{(C)}}\bigg\vert_{\{\br\} = \{\bR\}} 
\end{align}
\end{subequations}
and the first derivative terms (\ref{firstPhi}) vanish by the assumption of expansion about the equilibrium positions.
In arriving at (\ref{elasCD}) use has been made of the invariance of the entire crystal under spatial translations, 
so that the expansion (\ref{Phigrad}) can depend only upon the displacement gradients
\begin{equation}
u_{i,j} \equiv \frac{\pa u_i}{\pa x_j}
\label{ugrad}
\end{equation}
rather than the displacement (\ref{udisp}) itself, and the continuum limit is taken by the replacement of the discrete
lattice displacements $u_i^{(A)}  = u_i (\bR^{(A)})$ by the continuous function of position $u_i(\bx)$ to be integrated over. 
Since summing over all lattice sites in (\ref{elasCD}) results in an extensive quantity, proportional to the total number 
of lattice sites and hence to the total volume $V$ of the crystal, the factor of the total volume $V$ is extracted explicitly 
from definitions of the elastic constants (\ref{elasCD}) to yield intensive quantities independent of the total volume.

Substituting (\ref{elasCD}) into (\ref{latHam}), and setting to zero the potential energy function at its
equilibrium value $\Ph\{\bR\} =0$, we may pass to the continuum limit to arrive at the elastic Hamiltonian
\begin{equation}
H = \int d^3 \bx \, \bigg\{ \sdfrac{1}{2}\, \rh \sum_{i} \dot u_i^2 + \sdfrac{1}{2}\sum_{ijkl} \elC_{ijkl}\, u_{i, j}\,u_{k,l} 
+ \sdfrac{1}{6}\!\sum_{ijklmn} \!\elD_{ijklmn}\, u_{i,j}\,u_{k,l}\,u_{m,n}\bigg\} + \dots 
\label{Ham}
\end{equation}
where
\begin{equation}
\rh = \frac{1}{V} \sum_A M^{(A)}
\end{equation}
is the average continuum mass density of the solid in equilibrium. Clearly the expansion in gradients leading to this continuum 
elastic Hamiltonian (\ref{Ham}) is valid if and only if all the dimensionless gradients are small
\begin{equation}
\left\vert u_{i, j} \right\vert \ll 1
\label{strainsmall}
\end{equation}
{\it and} the spatial length scale of these gradients are very much greater than the lattice spacing scale of the underlying crystalline lattice \cite{Wallace:1972,Landau:1986,Kosevich:2005}.
These two restrictions should be kept in mind when considering the applicability of continuum elasticity theory to dislocations
and the drag coefficient due to interaction with phonons.

Variation of the continuum Hamiltonian (\ref{Ham}) with respect to $u_i (\bx)$ yields the equation of motion 
for the continuous displacement field
\begin{equation}
\rh \frac{\pa^2 u_i}{\pa t^2} = \sum_{jkl} \elC_{ijkl} \frac{\pa^2 u_k}{\pa x_j\pa x_l} + 
\sum_{jklmn} \elD_{ijklmn}\, \frac{\pa}{\pa x_j}\left(\frac{\pa u_k}{\pa x_l}\frac{\pa u_m}{\pa x_n}\right) + \dots
\label{eom} 
\end{equation}
with the harmonic approximation of elastic displacements and phonons obtained by neglect of all terms beyond the linear ones in (\ref{eom}). The next-order 
anharmonic terms involving $\elD_{ijklmn}$ act as a small perturbation to the linear approximation, and will describe the interaction of the phonons
with the moving dislocation field.

We shall also find it useful to introduce the notations
\begin{subequations}
\begin{align}
u_{(i,j)} &\equiv \sdfrac{1}{2} (u_{i,j} + u_{j,i}) \equiv \e_{ij} \\
u_{[i,j]} &\equiv\sdfrac{1}{2} (u_{i,j} - u_{j,i}) \equiv \w_{ij}
\end{align}
\label{swdef}\end{subequations}
for the symmetric and antisymmetric parts respectively of the displacement gradient (\ref{ugrad}). Although in most treatments 
of continuum elasticity the invariance of the solid to global rotations is used to eliminate all dependence upon the antisymmetric $\w_{ij}$, 
and indeed for the stress-strain relations of a perfect crystal only the symmetrized gradients $\e_{ij}$ can appear, both small deformations 
of the crystal lattice, as well as  the relatively larger dislocation defects, generally contain local regions with {\it both} symmetric and 
antisymmetric gradients, with non-zero rotation and torsional twistings of the solid, which should be taken into account. Since we are interested 
in this work on the interaction of defects with phonons, both of which may have antisymmetric strain fields, we retain $\w_{ij}$. As we shall see,
this generalization leads to new anharmonic interaction terms in (\ref{V3}) below that apparently have not been considered previously in the literature.

\subsection{Second-Order Elastic Constants and Phonons}

Although bulk continuum quantities, the elastic tensors $\elC_{ijkl}$ and $\elD_{ijklmn}$ still depend upon the underlying 
discrete symmetry group of the crystal, with many independent components \cite{Teodosiu:1982}.
A reasonable first approximation to the dislocation
dynamics is obtained by averaging over all directions and assuming that the undeformed crystal can be treated as homogeneous 
and isotropic. In that case the number of independent components that can appear in $\elC_{ijkl}$ and $\elD_{ijklmn}$ is greatly reduced. 
These independent components can be determined by consideration of the rotationally invariant scalars that
can be constructed from the displacement gradients \cite{Landau:1986,Wallace:1972}.

At second order in gradients of $\bf u$ there are only two possible scalars that can be constructed, namely
\begin{equation} 
(\na \cdot \bu)^2 = \big(u_{i,i}\big)^2 \qquad {\rm and} \qquad (\na \times \bu)^2 = 2\, \w_{ij}\,\w_{ij}  = \big(u_{i,j}\big)^2 - u_{i,j}u_{j,i}
\label{secinv}
\end{equation}
involving only the symmetric and anti-symmetric parts of the displacement gradient respectively. Thus the 
potential term in the continuum Hamiltonian (\ref{Ham}) may be expressed as the sum of two terms~\cite{Landau:1986}
\begin{equation}
\Phi_2 = \sdfrac{1}{2}\, (\l + 2 \m)\int d^3 \bx\, (\na \cdot \bu)^2 + \sdfrac{1}{2}\, \m\int d^3 \bx\, (\na \times \bu)^2
\label{secpot}
\end{equation}
at second order in the displacement gradients $u_{i,j}$. Up to an integration by parts and surface contribution which we neglect,
this corresponds to reducing the second order elastic tensor $C_{ijkl}$ to the sum of just two terms
\begin{equation}
\elC_{ijkl} = \l\,  \d_{ij} \d_{kl} + \m\, (\d_{ik}\d_{jl} + \d_{il} \d_{jk}) 
\label{Cten}
\end{equation}
governed by the two elastic constants (Lam\'e constants) with $\m$ the shear modulus.
Both $\l$ and $\m$ must be positive to insure stability.
Substituting in (\ref{eom}), this leads to the linearized continuum eqs. of motion \cite{Landau:1986,Hirth:1982}
\begin{align}
\rh \,\frac{\pa^2\bu}{\,\pa t^2}&=\m\,\na^2\bu+(\l+\m)\,\nabla\big(\nabla\cdot\bu\,\big)
\label{linelas}
\end{align}
for the displacement vector field $\bu (\bx, t)$, where for the moment we neglect the higher order anharmonic terms.

The antisymmetric strain $\w_{ij}$ appears naturally through the curl $\na \times \bu$, expressing
a non-zero local rotational distortion of the solid. Recalling that any vector $\bf u$ can be separated into its transverse (T) 
and longitudinal (L) parts by 
\begin{equation}
u_i= u_i^\txt{(T)} + u_i^\txt{(L)} = \left(u_i- \pa_i \frac{1}{\na^2}\pa_j u_j \right)  +  \pa_i \frac{1}{\na^2}\pa_j u_j 
\label{udecompose}
\end{equation}
the first with vanishing divergence, $\na\cdot \bu^\txt{(T)} = 0$ and the second with vanishing curl $\na \times \bu^\txt{(L)} = 0$,
we see from the definitions (\ref{swdef}) that the transverse (T) displacement vector can be expressed in terms of the anti-symmetrized 
strain $\w_{ij}$ (or vice versa) by
\begin{equation}
u_i^\txt{(T)} =  \frac{2\ }{\na^2}\, \pa_j \w_{ji}\,, \qquad\qquad \w_{ij} = \frac12 \vare_{ijk}\, (\na \times  \bu^\txt{(T)})_k 
= \frac12\vare_{ijk} (\na \times  \bu)_k
\label{wtr}
\end{equation}
while the longitudinal (L) displacement vector can be expressed in terms of the symmetrized strain $\e_{ij}$ (or vice versa) by
\begin{equation}
u_i^\txt{(L)} = \pa_i \frac{1\,}{\na^2} \e_{jj}\,, \qquad\qquad \e_{ii} = \pa_iu_i^\txt{(L)} = \na \cdot \bu
\label{slo}
\end{equation}
where we adopt the usual convention that repeated indices are to be summed, and $(\na^{2})^{-1}$ denotes the Green's function
inverse of the Laplacian.

The transverse and longitudinal components of $\bu$ are linearly independent, and the linearized continuum eq. (\ref{linelas}) 
separates into two independent linear equations
\begin{subequations}\label{linelasTL}
\begin{align}
\rh \,\frac{\pa^2\bu^\txt{(T)}}{\,\pa t^2}& = \m\, \na^2\bu^\txt{(T)}\\
\rh \,\frac{\pa^2\bu^\txt{(L)}}{\,\pa t^2}& = (\l+2 \m)\,\nabla\big(\nabla\cdot\bu^\txt{(L)}\,\big)
\end{align}
\end{subequations}
for the transverse and longitudinal displacements respectively.
Each of these equations has the form of a wave equation, which because of the relations (\ref{wtr}) and (\ref{slo}) may
be regarded as describing the propagation of anti-symmetric and symmetric strain fields through the solid medium,
but with a different speed of propagation, depending upon whether the displacement is transverse or longitudinal (\emph{i.e.}
perpendicular or parallel) to its direction of propagation.

The transverse and longitudinal wave solutions of (\ref{linelas}) are easily found in Fourier space.
The two transverse modes are 
\begin{align}
\bu^\txt{(T)}(\bx, t \vert \bk,s)&= \bfe(\bk, s) \, e^{i\bk\cdot \bx - i \wt t}\,, &&
\bk \cdot \bfe (\bk, s) = 0\,,&& s=1,2
\end{align}
with dispersion relation and sound speed
\begin{align}
\wt(k) &= \ct k \,, &&
\ct = \sqrt{\frac{\m}{\rh}}\,,
\end{align}
where $k \equiv |\bk|$ and $s = 1,2$ labels the two unit polarization vectors perpendicular to the direction of propagation $\hbk$.
The single longitudinal mode
\begin{align}
\bu^\txt{(L)}(\bx, t \vert \bk) &= \bfe(\bk, 3) \, e^{i\bk\cdot \bx - i \wl t}\,, && \bk\times \bfe(\bk, 3) = 0
\end{align}
with displacements in the direction of propagation, has dispersion relation and sound speed
\begin{align}
\wl(k) &= \cl k\,, && \cl = \sqrt{\frac{\l + 2 \m}{\rh}} > \ct 
\end{align}
where to simplify our subsequent algebra, we have employed the notation in which the polarization indices $s,s' =1,2$ label 
the two transverse sound modes, and $s, s'=3$ labels the single longitudinal mode. Thus the polarization vectors 
may be chosen to satisfy
\begin{equation}
\bfe(\bk,s)\cdot\bfe(\bk,s')= \d_{ss'}\qquad {\rm and} \qquad \sum_{s =1,2,3} \bfe_i(\bk,s)\,\bfe_j(\bk,s)= \d_{ij}
\label{eq:pol}
\end{equation}
and form an orthonormal basis set.

Upon quantization~\cite{Kosevich:2005}, the small elastic displacement fields in the continuum limit may be represented in terms of three independent phonon modes by 
\begin{align}
\bu(\bx, t)\Big\vert_{\txt{phonon}}\!\!=  \sqrt{\frac{\hbar}{\rh}} 
\sum_s\! \intk\!\! \frac{1}{\sqrt{2 \w_s(k)}} \left\{\ha_s(\bk)e^{i \bk \cdot \bx - i \w_{s}t } + \ha_s^{\dagger}(\bk)e^{-i \bk \cdot \bx + i \w_{s}t } \right\} \bfe(\bk,s)
\label{eq:phononmodes}
\end{align}
where the phonon creation and destruction operators are quantized with the continuum normalization
\begin{equation}
[\ha_{s}(\bk), \ha^{\dagger}_{s'}(\bk')] = (2 \pi)^3 \, \d^3(\bk -\bk')\, \d_{s s'}
\label{eq:phononcommutator}
\end{equation}
in the infinite volume limit. At linear order of continuum elastic theory in the isotropic crystal approximation we are using,
the three phonon modes all have gapless linear dispersion relations $\wt(k) = \ct k$, $\wl(k) = \cl k$, characteristic 
of the Debye approximation.

\subsection{General Third-Order Continuum Elastic Constants for an Isotropic Solid}
\label{sec:elastic-constants}

In the bulk of the existing literature, the continuum elastic potential has been taken to be dependent upon only the symmetrized
strain field $\e_{ij}$. This is possible up to quadratic order term $\Phi_2$ because the volume integral of the second (curl) term in (\ref{secpot}) is
\begin{eqnarray}
&&\hspace{-7mm}\sdfrac{\m}{2}\int d^3 \bx\, (\na \times \bu)^2 = \m\! \int \!d^3 \bx \, \w_{ij}\w_{ij} = \sdfrac{\m}{2}\!\int\! d^3 \bx \left(u_{i,j}u_{i,j} - u_{i,j}u_{j,i}\right) 
= \sdfrac{\m}{2}\!\int \!d^3 \bx \left(u_{i,j}^2 - u_{i,i}u_{j,j}\right) \nn\\
&&\hspace{2.4cm} =\m \!\int\!d^3 \bx \big[\e_{ij}\e_{ij} - (\e_{ii})^2\big]
\label{wsdegen}
\end{eqnarray}
after freely integrating by parts and dropping surface terms. Thus the contribution of the anti-symmetric strain field contribution $\w_{ij}$
proportional to $\na \times \bu$ may be expressed in terms of the symmetrized strain field at quadratic order for an isotropic solid. 
This is consistent with the literature, {\it e.g.} Ref. \cite{Murnaghan:1937,Murnaghan:1951}, in which the two terms in the elastic energy for an isotropic solid
at second order in the strain field are written
\be
\Phi_2 = \int\!d^3 \bx \left\{ \frac{\left(\l + 2 \m\right)}{2}\, I_1^2 - 2 \m I_2\right\}
\label{Phi2Murn}
\ee
in terms of the fundamental invariants
\be
I_1 \equiv {\rm Tr}(\e) = \e_{ii}\,, \qquad {\rm and} \qquad I_2 \equiv {\rm Tr (co}[\e]) = -\sdfrac{1}{2} \left[\e_{ij}\e_{ij}-  (\e_{ii})^2 \right]
\label{I1I2}
\ee
composed only of the symmetric strain field matrix $\e_{ij}$. Here co$[\e]$ denotes the cofactor matrix of $\e_{ij}$, with matrix elements
$({\rm co} [\e])_{ij} = \vare_{ikl}\vare_{jmn} \e_{km}\e_{ln}/2$. 

After the integration by parts needed to establish (\ref{wsdegen}), (\ref{Phi2Murn}) with (\ref{I1I2}) therefore coincides with (\ref{secpot}). However, the fundamental degrees of freedom reside in the displacement vector field $\bu$, and both its decomposition (\ref{udecompose}) and its equation of motion (\ref{linelas}) show that transverse modes with non-zero asymmetric rotational strains propagate. Moreover, these anti-symmetric strains cannot 
be eliminated from the third order terms by integrations by parts analogous to (\ref{wsdegen}), and lead to novel interactions 
between phonons and dislocations, as we show next.

If we allow for the invariant contractions of the anti-symmetric strain field $\w_{ij}$ to appear in (\ref{Ham}) we obtain the general third-order elastic potential
\begin{align}
\Phi_3 &= \!\!\int\!\! d^3\bx \Big\{a\,(\e_{ii})^3\! + b\,\e_{ii} \e_{jk}\e_{jk} + c\,\e_{ij}\e_{jk} \e_{ki} + g\,\e_{ii}\w_{jk}\w_{jk} 
+ h\,\e_{ij}(\vare_{ikl}\w_{kl})(\vare_{jmn} \w_{mn}) \Big\}\nonumber\\
&= \int d^3\bx \left\{a\, (u_{i,i})^3 + \left(\sdfrac{b}{2} + \sdfrac{g}{2} + h\right) u_{i,i}\, u_{j,k}\,u_{j,k} 
+ \left(\sdfrac{b}{2} - \sdfrac{g}{2} - h\right) u_{i,i}\, u_{j,k}\,u_{k,j} \right.\nonumber\\
& \hspace{2cm}\left. + \bigg(\sdfrac{c}{4} + h\bigg) u_{i,j}\, u_{j,k}\,u_{k,i} +\left(\sdfrac{3c}{4} - h\right) u_{i,j}\, u_{i,k}\,u_{j,k}\right\}
\label{V3}
\end{align}
where the $a,b,c$ coefficients multiply terms which are totally symmetric, while the $g,h$ coefficients multiply terms with one symmetric (S), 
and two antisymmetric (A) gradients, which we denote as mixed SAA terms. Terms with an odd number of $\w_{ij}$ vanish identically.
The general third-order potential for an isotropic solid corresponds to the third-order elastic tensor
\begin{align}
&\elD_{ijklmn} = 6a \,\d_{ij}\d_{kl}\d_{mn}
+ (b + g +2h)\Big(\d_{ij}\d_{km}\d_{ln}+\d_{im}\d_{jn}\d_{kl}+\d_{ik}\d_{jl}\d_{mn}\Big)\nonumber\\
&+ (b - g -2h)\Big(\d_{ij}\d_{kn}\d_{lm}+\d_{in}\d_{jm}\d_{kl}+\d_{il}\d_{jk}\d_{mn}\Big)
+ 3 \left(\sdfrac{c}{4} + h\right )  \Big( \d_{il}\d_{jm}\d_{kn}+  \d_{in}\d_{jk}\d_{lm}\Big)\nonumber\\
&+\left(\sdfrac{3c}{4}- h\right)\Big[\d_{ik}\left(\d_{jm}\d_{ln}+\d_{jn}\d_{lm}\right) + \d_{im}\left(\d_{jl}\d_{kn}+\d_{jk}\d_{ln}\right)
+ \d_{km} \left(\d_{in}\d_{jl} + \d_{il}\d_{jn}\right)\Big]
\label{V3t}
\end{align}
which as defined in (\ref{Ham}) is symmetric under interchange of any of the three pairs of indices $(i,j), (k,l), (m,n)$.
Note that at this third order the $g$ and $h$ terms involving the anti-symmetrized strains $\w_{ij}$ cannot be eliminated or expressed in terms of the symmetrized strains by integration 
by parts as in (\ref{wsdegen}). All {\it five} coefficients of the third-order terms in (\ref{V3})-(\ref{V3t}) are independent of each other, and in general 
{\it all five} coefficients  take on non-zero values for perturbations from equilibrium, either by phonons or dislocations. 
Indeed, \eqref{V3t} generalizes earlier results:

It appears that the independence of the third-order mixed elastic constants $g,h$ from the $a,b,c$ and $\l, \m$ elastic constants 
has not been fully taken into account in the literature due to the fact that only the symmetrized strain field $\e_{ij}$ has been considered 
relevant, and the independent $(g,h)$ SAA terms in the third-order elastic energy (\ref{V3}) have been neglected, perhaps in part because they
are difficult to measure independently. For example, Murnaghan defines the three constants $(\ml$, $\mm$, $\mn)$ 
by the third-order elastic energy \cite{Murnaghan:1937,Murnaghan:1951}
\be
\tilde\Phi_3= \int d^2\bx \left\{\frac{(\ml + 2 \mm )}{3} \, I_1^3  - 2 \mm\, I_1I_2 + \mn\, I_3\right\}
\label{Phi3Murn}
\ee
for an isotropic solid, in terms of the invariants $I_1,I_2$ defined in (\ref{I1I2}), and
\be
I_3 \equiv {\rm det}(\e) = \sdfrac{1}{6}\, (\e_{ii})^3 - \sdfrac{1}{2}\,\e_{ii}\, \e_{jk}\,\e_{jk} + \sdfrac{1}{3}\, \e_{ij}\,\e_{jk}\,\e_{ki}
\ee
composed of the symmetric strain matrix $\e_{ij}$ only. Likewise Refs.~\cite{Landau:1986} and \cite{Toupin:1961} give the 
same general expression (\ref{V3t}) for the isotropic third-order elastic tensor as Ref.~\cite{Murnaghan:1937,Murnaghan:1951}, which is equivalent to
\begin{align}
\tilde \elD_{ijklmn} &= \elD_{ijklmn}(a,b, c \rightarrow \tilde a, \tilde b, \tilde c) \Big\vert_{g=h=0} 
\label{tildeD}
\end{align}
with the SAA $g,h$ terms set to zero, in terms of three constants $(A, B, C)$ or $(\n_1,\n_2, \n_3)$ respectively,
which are related to $(\tilde a, \tilde b, \tilde c)$ and Murnaghan's constants $(\ml,\mm,\mn)$ in (\ref{Phi3Murn}) by \cite{Volkov:2015}
\begin{align}
\tilde a &= \sdfrac{\n_1}{6} = \sdfrac{C}{3}= \sdfrac{1}{3}\left(\ml - \mm\right) + \sdfrac{ \mn}{6} &
\tilde b &= \n_2 = B= \mm - \sdfrac{\mn}{2} &
\tilde c &= \sdfrac{4}{3}\, \n_3 = \sdfrac{A}{3} = \sdfrac{\mn}{4}
\label{eq:relationtoMurnaghan}
\end{align}
with $g=h=0$, as in (\ref{tildeD}).

On the other hand it has also been common practice to expand the crystal potential in the non-linear strain field 
(known as the Murnaghan strain~\cite{Murnaghan:1937,Murnaghan:1951} or Green-Saint-Venant strain tensor~\cite{Lubliner:2008})
\begin{equation}
\eta_{ij} \equiv \frac{1}{2} \left(\frac{\pa u_i}{\pa x_j} +  \frac{\pa u_j}{\pa x_i} +  \frac{\pa u_k}{\pa x_i}  \frac{\pa u_k}{\pa x_j} \right) 
= \e_{ij} + \sdfrac{1}{2} u_{k,i} \,u_{k,j} 
\label{secstrain}
\end{equation}
which is fully symmetric in indices $i,j$, but also contains a non-linear term quadratic in the strains. Then if the two independent SAA constants, 
$g$ and $h$, are set to zero in $\Phi_3$, but the total potential is taken to be
\begin{equation}
\sdfrac{1}{2} \tilde \elC_{ijkl} \eta_{ij}\eta_{kl} + \sdfrac{1}{6} \tilde \elD_{ijklmn}\eta_{ij} \eta_{kl} \eta_{mn}
\label{symthird}
\end{equation}
in terms of $\eta_{ij}$, the second-order tensor
\begin{equation}
\tilde C_{ijkl} = C_{ijkl}
\end{equation}
is identical to (\ref{Cten}) for an isotropic crystal, but the elastic energy to third order in the linearized strains $u_{i,j}$ receives contributions from 
the second-order Lam\'e coefficients $\l$ and $\m$ due to the non-linear quadratic term in (\ref{secstrain}). Taking the total potential to be (\ref{symthird}), 
and expanding consistently up to third order in the strains $u_{i,j}$, is equivalent to the \emph{effective} third-order elastic tensor \cite{Wallace:1970,Wallace:1972}
\begin{align}
D_{ijklmn}\Big\vert_{eff}&= \tilde \elD_{ijklmn}+  C_{ijln}\d_{km}+ C_{kljn}\d_{im}+C_{jlmn}\d_{ik}
 \,, \label{AinC}
\end{align}
in our expressions. This leads to the same $\Phi_3$ as (\ref{V3}), and all five tensor structures, but with
\begin{subequations}\label{eq:constantsabcgh}
\begin{align}
a_{eff}&= \tilde a \,, &
b_{eff}& = \tilde b +\sdfrac{1}{2}\,\l \,, &
c_{eff}& = \tilde c+ \m \,, \\
g_{eff}& = \sdfrac{1}{2}\left(\l + \m\right)\,, &
h_{eff}& = -\sdfrac{1}{4}\,\m
\,,
\end{align}
\end{subequations}
and the $g$ and $h$ terms effectively generated by the lower-order Lam\'e coefficients $\l,\m$, instead of being truly independent 
third-order constants as they are in the more general expression (\ref{V3}), in which the anti-symmetric strains $\w_{ij}$ are allowed
from the outset. The additional independent SAA terms with general $g,h$ are relevant for dislocations interacting with phonons, 
and hence the phonon wind drag coefficient, and ideally should be independently measured.

Because of the intrinsic uncertainties in applying the isotropic approximation to metals of interest with definite crystalline symmetries~\cite{Lubarda:1997,Blaschke:2017Poly},
and the fact that the independent third order constants $g$ and $h$ are not available for these metals, we will use the experimentally measured 
effective isotropic constants of Table~\ref{tab:values-metals} for our study of the velocity dependence of the drag coefficient in Section~\ref{sec:results}. 
The more general theory developed here may be adapted and refined when accurate values of the five independent third order elastic constants become available
for these metals.

\section{Moving Edge and Screw Dislocations}
\label{sec:dislocations}

A stationary infinite edge dislocation along the $\bf\hat z$ axis with Burgers vector in the $\bf\hat x$ direction is described by the displacement vector
$\bU (\bfx)$ with components \cite{Burgers:1939a,Weertman:1961}
\begin{subequations}
\label{edgestatic}
\begin{align}
U_x (x,y)\Big\vert_{edge} &= \frac{b}{2 \pi} \left\{ \tan^{-1}\left(\frac{y}{x}\right) +  \left(\frac{\l + \m}{\l + 2 \m}\right)\frac{xy}{x^2 + y^2}\right\}\\
U_y (x,y)\Big\vert_{edge}  &= \frac{b}{2 \pi}\left(\frac{1}{\l + 2 \m} \right) \left\{ -\frac{\m}{2} \ln\left(\frac{x^2 + y^2}{r_0^2}\right) +   \left(\l + \m\right)\frac{y^2}{x^2 + y^2}\right\}\\
U_z\Big\vert_{edge} &= 0
\end{align}
\end{subequations}
where $b$ is the magnitude of the Burgers' vector, and $r_0$ is the dislocation core radius. We denote the finite displacement vector of a dislocation by
a capital $\bU$ to distinguish it from the linearized small displacement $\bu$ of elasticity theory in the previous section.

Similarly a stationary infinite screw dislocation along the $\bf\hat z$ axis with Burgers vector ${\bf b} = b {\bf \hat z}$ is described by the displacement vector $\bU (\bfx)$ with components \cite{Burgers:1939a,Weertman:1961}
\begin{subequations}\label{screwstatic}
\begin{align}
U_x\Big\vert_{screw}  &= U_y\Big\vert_{screw}   = 0\\
U_z(x,y)\Big\vert_{screw}  & = \frac{b}{2 \pi} \, \tan^{-1}\left(\frac{y}{x}\right) \, .
\end{align}
\end{subequations}
Each of these dislocation displacements are solutions of the equations of linearized static continuum elasticity (\ref{linelas}), everywhere except 
at the origin $x=y=0$, where they are singular.
The edge dislocation (\ref{edgestatic}) contains both longitudinal (L) components with zero curl and transverse components (T) with zero divergence, 
whereas the screw dislocation (\ref{screwstatic}) is purely transverse (T). The non-zero transverse displacements have anti-symmetric rotation strain fields, 
hence $\w_{ij} = u_{[i,j]} \neq 0$, and both dislocations have non-zero internal torques. This makes non-zero contributions to SAA terms in the third order
elastic energy (\ref{V3}).

In order to find the solutions of the linear elastic eqs. (\ref{linelas}) for edge dislocations moving with uniform velocity ${\bf v} = v \bf \hat x$
(with $\mathbf{\hat\e}=\mathbf{\hat z}$, $\mathbf{\hat b}=\mathbf{\hat x}$), Eshelby decomposed the stationary edge dislocation (\ref{edgestatic}) 
into its transverse and longitudinal parts as in (\ref{udecompose}), using the effective Lorentz invariance of the eqs. (\ref{linelasTL}), 
and obtained \cite{Eshelby:1949}
\begin{align}
\hspace{-3mm}U_x (x,y;t)\Big\vert_{edge}  \hspace{-2mm}&=
\frac{b}{\pi\bt ^2}\left\{\tan^{-1}\!\left[\frac{y}{\gl (x-vt)}\right] -\left(1-\sdfrac{\bt ^2}{2}\right)\tan^{-1}\!\left[\frac{y}{\gt (x-vt)}\right]\right\}\nn\\
\hspace{-3mm}U_y (x,y;t)\Big\vert_{edge}  \hspace{-2mm}&=
\frac{b}{2\pi\bt ^2}\!\left\{\!\frac1{\gl }\ln\!\left[\frac{(x-vt)^2+y^2/\gl ^2}{r_0^2}\right] -{\gt}\!\left(1-\sdfrac{\bt ^2}{2}\right)\ln\!\left[\frac{(x-vt)^2+y^2/\gt ^2}{r_0^2}\right]\right\} 
\label{eq:dxy-regul}
\end{align}
for an edge dislocation gliding in an isotropic elastic solid in the $x$ direction with uniform velocity $v$, where
\begin{equation}
\b_{_{T,L}} \equiv \frac{v}{c_{_{T,L}}}\,,\qquad  \g_{_{T,L}}= \frac{1}{\sqrt{1-\b_{_{T,L}}^2}}
\end{equation}
following the standard notations of special relativity, whose definition of $\g$ is related to that of Eshelby by $\g\to1/\g$.
We consider below only gliding edge and screw dislocations, as dislocation climb is highly suppressed
and hence can be neglected in the discussion of phonon wind~\cite{Alshits:1992}.


One may check that $\bU (x,y;t)$ above satisfies \eqnref{linelas}. In fact, the first (resp. second) terms of $\bU (x,y;t)$ depending only on $\gl $ (resp. $\gt $) 
satisfy \eqnref{linelasTL} {\it independently} of each other. However, only with this particular combination in \eqref{eq:dxy-regul} will no external concentrated 
force need to be applied in the $y$-direction at the core of the dislocation (where $x=y=0$); see e.g.~\cite{Weertman:1980}. We note that $U_y$ differs from 
some other results found in the literature by an arbitrary constant. For example, if we take the limit $v\to0$ in \eqref{eq:dxy-regul}, and express the results 
in terms of the Poisson ratio $\n=\l/2(\m+\l)$, our expression for $U_y$ differs from those in Ref.~\cite{Hirth:1982} by a constant ${b}/{8\pi(1-\n)}$, 
but agrees with the original result of Burgers~\cite{Burgers:1939a}. Since the interaction Hamiltonian \eqref{HintA} depends only on the gradient of 
the dislocation displacement, this additive constant is of no physical relevance.



\begin{figure}[h!t!b]
\centering
 \vspace*{-0.9cm}
 \includegraphics[width=\textwidth]{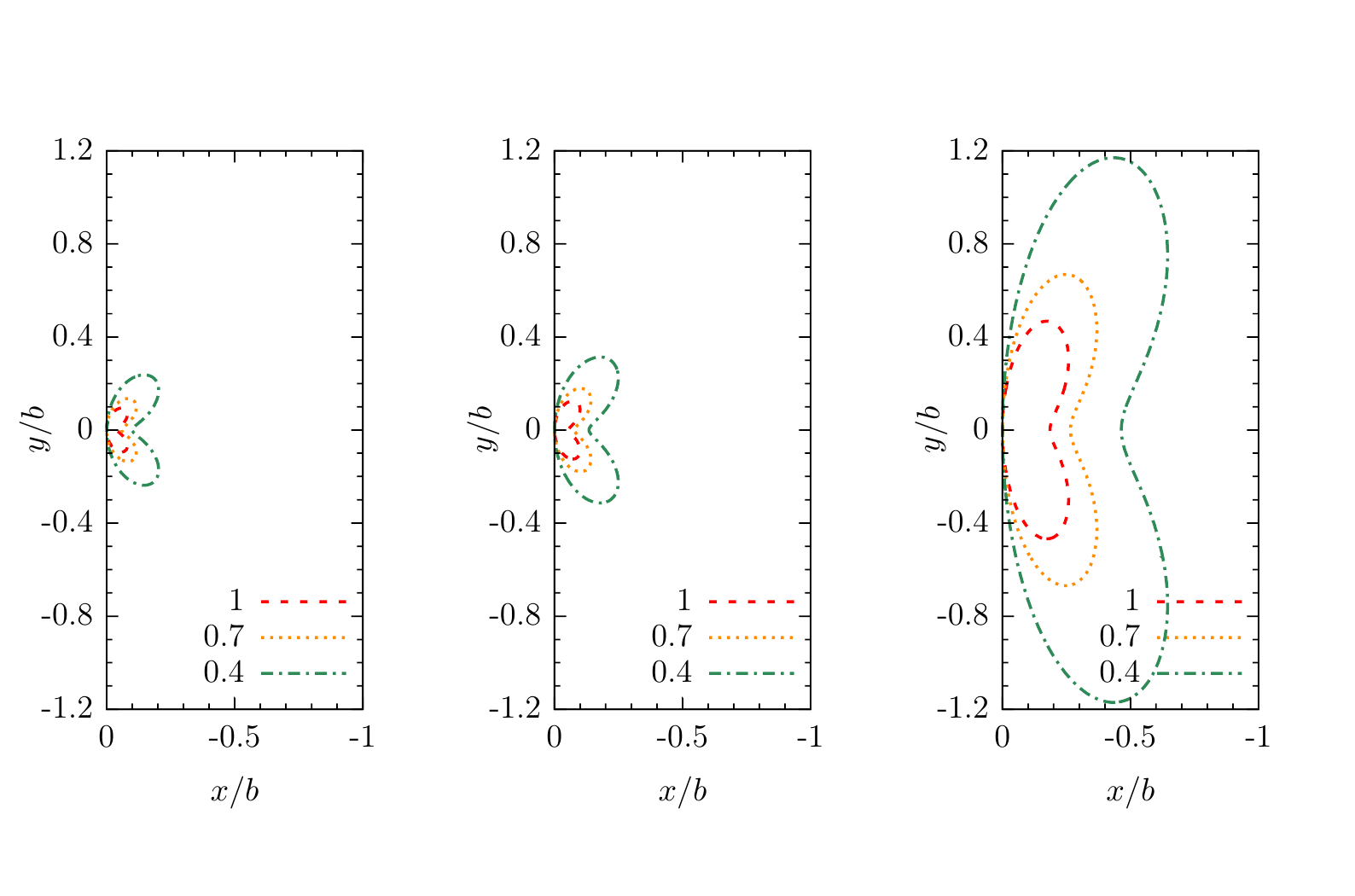}
 \vspace*{-1.4cm}
\caption{Level curves of the gradient $U_{y,x}$ for edge dislocations are shown (from left to right) for velocities $v=0$, $v=0.5\ct$, and $v=0.9\ct$. The gradient at $(x,y)$ increases with $v$, hence there is an expansion of the core region where the assumption of linear elasticity, $U_{ij}\ll 1$, breaks down.  Note that the core expansion is not uniform: the elongation along the $y$-axis is greater than along the $x$-axis.}
\label{fig:contour-edge}
\end{figure}

{\allowdisplaybreaks
The gradients of the moving edge dislocation displacement field \eqref{eq:dxy-regul} are
\begin{subequations}
\label{edgedis}
\begin{align}
U_{x,x}\Big\vert_{edge}&=-\frac{b y}{\pi  \bt ^2 }\left[\frac{1}{\Big(\gl (x-vt)^2 +y^2/\gl\Big)}-\frac{1-\bt^2/2}{\Big(\gt(x-vt)^2 +y^2/\gt\Big)}\right]\\
U_{x,y}\Big\vert_{edge}&=\frac{b (x-vt)}{\pi  \bt ^2 }\left[\frac{1}{\Big(\gl (x-vt)^2 +y^2/\gl\Big)}-\frac{1-\bt^2/2}{\Big(\gt(x-vt)^2 +y^2/\gt\Big)}\right]\\
U_{y,x}\Big\vert_{edge}&=\frac{b (x-vt)}{\pi  \bt ^2 }\left[ \frac{1}{\Big(\gl (x-vt)^2 +y^2/\gl\Big)}-\frac{\gt^2\left(1-\bt^2/2\right)}{\Big(\gt(x-vt)^2 +y^2/\gt\Big)}\right]\\
U_{y,y}\Big\vert_{edge}&=\frac{b y}{\pi  \bt ^2 }\left[ \frac{1/\gl^2}{\Big(\gl (x-vt)^2 +y^2/\gl\Big)}-\frac{1-\bt^2/2}{\Big(\gt(x-vt)^2 +y^2/\gt\Big)}\right]\,.
\end{align}
\end{subequations}
Note that the short-distance cutoff $r_0$ of (\ref{edgestatic}) does not appear in the gradient strains (\ref{edgedis}), thus the energy of the edge
dislocation field does not depend on the core cutoff. Nevertheless, we must anticipate that the continuum approximation will break
down when the strain fields are large in the core region where the inequality (\ref{strainsmall}) for $U_{i,j}$ is violated. This condition
is violated at velocities approaching transverse sound speed, as illustrated in Fig.~\ref{fig:contour-edge}.
}

\begin{figure}[h!t!b]
\centering
 \vspace*{-0.9cm}
 \includegraphics[width=\textwidth]{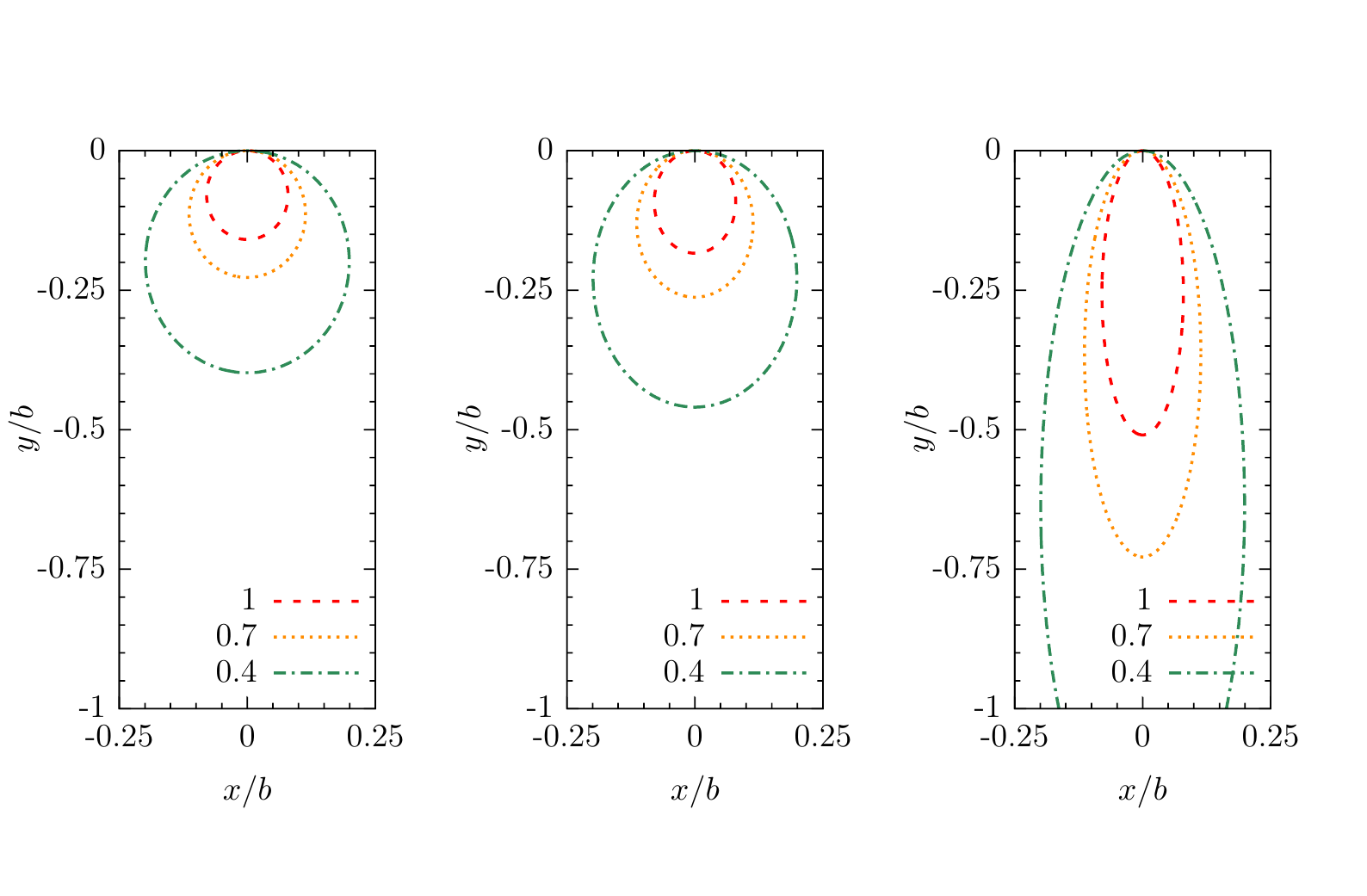}
 \vspace*{-1.4cm}
\caption{Level curves of the gradient $U_{z,x}$ for screw dislocations are shown (from left to right) for velocities $v=0$, $v=0.5\ct$, and $v=0.95\ct$.  As in the previous figure for edge dislocations, there is a non-uniform expansion of the core region.  Here the expansion occurs only along the $y$-axis.}
\label{fig:contour-screw}
\end{figure}

The time-dependent displacement field of a screw dislocation with sense vector $\hat\e$ along the positive $\bf \hat z$-axis, 
gliding in the $x$ direction at velocity $\bfv = v \bf{\hat x}$, is given by~\cite{Eshelby:1949}
\begin{align}
U_z(x,y;t)\Big\vert_{screw} &=\frac{b}{2\pi}\tan^{-1}\left[\frac{y}{\gt (x-vt)}\right]
\label{eq:dxy-screw}
\end{align}
and it may be checked that this displacement field $\bU$ with non-zero curl satisfies \eqnref{linelas}.
Its non-vanishing strain field gradient components are 
\begin{subequations}
\begin{align}
U_{z,x}(x,y;t)\Big\vert_{screw}&=-\frac{b}{2 \pi} \frac{y}{\big[\gt (x-vt)^2+y^2/\gt\big]}\,, \\
U_{z,y}(x,y;t)\Big\vert_{screw}&=\frac{b}{2 \pi} \frac{(x-vt)}{\big[\gt(x-vt)^2+y^2/\gt \big]} \, ;
\end{align}
\end{subequations}
the velocity-dependent contours of $U_{z,x}$ are illustrated in Fig.~\ref{fig:contour-screw}.

For the computation of the drag coefficient we will make use of the Fourier transforms of the dislocation fields $\tilde U_{i,j} (\bfq)$, defined by
\begin{align}
U_{i,j} (\bfx, t) &= \intq \,\tilde U_{i,j} (\bfq)\, e^{i \bfq\cdot(\bfx - \bfv t)}
\end{align}
with the time dependence simply taken into account by the $e^{-i\bfq\cdot \bfv t}$ phase factor. Expressing the momentum variable
$\bf q$ in cylindrical $(q, \phi, q_z)$ coordinates and making use of the independence of (\ref{edgedis}) of $z$ to write
\be
\tilde U_{i,j} (\bfq) = 2 \pi \d (q_z)\, \tilde U_{i,j} (q, \phi)
\ee
we find
\begin{subequations}
\begin{align}
\tilde U_{x,x}(q,\phi)\Big\vert_{edge}&=\frac{2 i b \sin \ph}{\bt ^2 q}
\left[\frac{1}{\left(1-\bl ^2 \cos ^2\ph\right)}-\frac{1-\bt^2/2}{\left(1-\bt ^2 \cos ^2\ph\right)}\right]\\
\tilde U_{x,y}(q,\phi)\Big\vert_{edge}&=\frac{-2 i b \cos \ph}{\bt ^2 q }
\left[\frac{1}{\gl^{2} \left(1-\bl ^2 \cos ^2\ph\right)}-\frac{1-\bt^2/2}{\gt^2\left(1-\bt ^2 \cos ^2\ph\right)}\right]\\
\tilde U_{y,x}(q,\phi)\Big\vert_{edge}&=\frac{-2 i b \cos \ph}{\bt ^2 q }
\left[\frac{1}{\gl^2\left(1-\bl ^2 \cos ^2\ph\right)}-\frac{1-\bt^2/2 }{\left(1-\bt ^2 \cos ^2\ph\right)}\right]\\
\tilde U_{y,y}(q,\phi)\Big\vert_{edge}&=\frac{-2 i b  \sin \ph}{\bt ^2 q }
\left[\frac{1}{\gl^2\left(1-\bl ^2 \cos ^2\ph\right)}-\frac{1-\bt^2/2}{\left(1-\bt ^2 \cos ^2\ph\right)}\right]
\end{align}
\label{eq:strains-nocutoffs}
\end{subequations}
for the edge dislocation. 

Likewise, we have
\begin{align}
\tilde U_{z,x}(q,\phi)\Big\vert_{screw}&=\frac{i b \sin\phi}{q \left(1-\bt ^2 \cos^2\phi\right)}\,,&
\tilde U_{z,y}(q,\phi)\Big\vert_{screw}&=-\frac{i b\, \cos\phi}{q\,\gt^2 \left(1-\bt ^2 \cos^2\phi\right)} 
\label{eq:strains-screws-nocutoffs}
\end{align}
for the Fourier transform of the corresponding strain field gradient for the screw dislocation, with all other components vanishing.

We note that the trace of the edge dislocation strain field simplifies to
\be
\tilde U_{i,i}(q,\phi)\Big\vert_{edge}= \frac{2 i b\, \ct^2 \sin \ph}{q\, \cl^2 \left(1-\bl ^2 \cos ^2\ph\right)}
\label{eq:trace-edge}
\ee
while the screw dislocation strain field has vanishing trace, depending as it does on only the transverse anti-symmetric 
components of the strain.

\section{The Phonon Wind Contribution to the Drag Coefficient}
\label{sec:phononwind}

The thermal phonons in a crystal are scattered by gliding dislocations, thereby resulting in a drag force on the dislocations.
By analogy with linear (Stokes) drag on objects moving through fluids at low velocities (low Reynolds numbers), the dislocation drag force per unit length is written
\begin{equation}
F = B \, v \, ,
\end{equation}
where $B = B(v, T)$ is the dislocation drag coefficient. At low velocities $B$ is approximately independent of the dislocation velocity, and hence the drag 
force $F$ is approximately linear in $v$. Typical low-velocity values of $B$ are in the range $10^{-4}$ -- $10^{-3}$ Poise ($10^{-2}$ -- $10^{-1}$ mPa-sec).
The energy dissipated per unit time per unit length of a dislocation is
\begin{equation}
D = F \, v = B \, v^2 \, .
\end{equation}

In discussing dislocations interacting with phonons, {\it i.e.} the phonon wind, the Hamiltonian of interest consists of a sum of two terms
\begin{align}
H&=H_{\txt{ph}} +H_\txt{int}(t) 
\end{align}
where $H_\txt{ph}$ is the free phonon contribution while $H_\txt{int}(t)$ is the interaction Hamiltonian between the phonons and 
the moving dislocation, which is time dependent.
In the following we will denote the dislocation wave vector by $\bfq$, whereas the phonon momenta 
will be denoted $\bfk$, $\bfki$. In the continuum limit where the lattice spacing goes to zero, the discrete sum over momenta may be replaced by an 
integral over the first Brillouin zone (BZ). Thus the phonon Hamiltonian is 
\begin{align}
H_\txt{ph}&=\frac{\hbar}{2}
\sum_{s} \intk \,\w_s(k)\left(\bfa^\dagger_{s}(\bfk)\,\bfa_{s}(\bfk)+\bfa_{s}(\bfk)\,\bfa^\dagger_{s}(\bfk)\right)
\label{eq:Hph}
\end{align}
where the quantized phonon mode creation and destruction operators obey the commutation relation \eqref{eq:phononcommutator}, 
in the continuous momentum variables, $\bfk$, $\bfki$.

The interaction Hamiltonian $H_\txt{int}$ may be obtained from \eqnref{V3} by reinterpreting the displacements as superpositions 
of phonons and displacements due to a moving dislocation. For the displacement gradients appearing in $\Phi_3$ this means replacing 
$u_{i,j}\to u_{i,j}+U_{i,j}$, and retaining the terms linear in the dislocation gradient $U_{i,j}$. This gives a trilinear interaction between 
two phonons and a single moving dislocation of the form
\begin{align}
H_\txt{int}(t) &= \frac{1}{2!}\sum_{ijklmn}\, D_{ijklmn}
  \intx \,  u_{i,j} (\bfx)\,u_{k,l}(\bfx) \,U_{m,n}(\bfx,t) \nn\\
 &=\int d^3\bx \bigg\{\left(\ml-\mm+\sdfrac{\mn}{2}\right) (u_{i,i})^2 U_{j,j} + \sdfrac{1}{2}\left(\mm-\sdfrac{\mn}{2}+\lambda\right) \left(u_{j,k}\,u_{j,k}\, U_{i,i}
 +2u_{i,i}\, u_{j,k}\,U_{j,k}\right) \nn\\
 &\qquad\qquad + \sdfrac{1}{2}\left(\mm-\sdfrac{\mn}{2}\right) \left(u_{j,k}\,u_{k,j}\, U_{i,i}+2u_{i,i}\, u_{j,k}\,U_{k,j}\right) 
 + \sdfrac{\mn}{4}\, u_{i,j}\, u_{j,k}\,U_{k,i} \nn\\
 &\qquad\qquad +\left(\sdfrac{\mn}{4} + \mu\right) \left(u_{i,k}\,u_{k,j}\,U_{i,j}  +u_{i,j}u_{k,j}\,U_{i,k}+u_{i,j}\, u_{i,k}\,U_{k,j}\right)\bigg\}
  \,.\label{HintA}
\end{align}
with elastic constants following from \eqref{V3}, \eqref{V3t}, \eqref{eq:relationtoMurnaghan}, and \eqref{eq:constantsabcgh}.
The elastic deformation fields are given in terms of the phonon modes \eqref{eq:phononmodes} by
\begin{align}
u_{i,j}(\bfx) &=i \,\sqrt{\frac{\hbar}{\rh }} \sum_s \intk \frac{\bfk_{j}}{\sqrt{2\w_s(k)}} \left(\bfa_s(\bfk) + \bfa_s^{\dagger}(-\bfk)\right)
e^{i \bfk \cdot \bfx} \, \bfe_{i} (\bfk, s)
\end{align}
in the continuum limit of an infinite crystal. In the following we will employ the definition
\begin{align}
\hat A_s(\bfk) &\equiv \bfa_s(\bfk) + \bfa_s^{\dagger}(-\bfk) = \hat A_s^{\dagger}(-\bfk)
\end{align}
in order to shorten lengthy expressions.

Substituting these relations into \eqref{HintA} and noting that the integration over $\bfx$ gives a momentum conserving
$\d$-function which sets $\bfq = \bfk- \bfki$ (upon changing the sign of $\bfk$ under the integral), we secure
\begin{align}
H_\txt{int}(t) &=  \frac12\sum_{ss'} \intk  \intki \,\hat A_s^{\dagger}(\bfk)\, \hat A_{s'} (\bfki)\, {\cal V}_{ss'}(\bfk,\bfki)
\,e^{-i(\bfk -\bfki)\cdot \bfv t}
\label{HintG}
\end{align}
where
\be
{\cal V}_{ss'}(\bfk,\bfki) = \frac{\hbar}{2\rh }\,\frac{1}{\!\!\sqrt{\w_s(k)\,\w_{s'}(k')}}
\sum_{ijklmn}\!\! \elA_{ijklmn}\,
\bfe^*_i(\bfk, s)\,\bfk_j\,  \bfe_k(\bfki, s') \,\bfki_l \,\tilde U_{mn}(\bfk-\bfki)
\label{eq:defGamma}
\ee
is the vertex describing the interaction between the dislocation and the phonon modes, and
$\tilde U_{i,j} (\bfq)$ denotes the Fourier transform of the deformation field of the dislocation.

Taking account of notational changes, the form of the vertex ${\cal V}_{ss'}(\bfk,\bfki) $ in \eqref{eq:defGamma} coincides with the one given 
in Ref.~\cite{Alshits:1979}. See also the earlier Refs.~\cite{Alshits:1969a,Alshits:1969b,Alshits:1973}. We note also that since 
the phonon wave vectors lie in the first Brillouin zone, $\abs{\bfk},\, \abs{\bfki}\le q_{_{BZ}}$, the dislocation wave vector magnitude
satisfies $q=\abs{\bfk-\bfki}\le 2q_{_{BZ}}$ due to momentum conservation.

Calculating the matrix elements of the interaction Hamiltonian (\ref{HintG}) between phonon states $(\bfk, s)$ and $(\bfki,s')$,
and defining $\W_q \equiv \abs{\bfq \cdot \bfv} = \abs{(\bfk-\bfki)\cdot \bfv}$, the probability per unit time of these transitions is \cite{Alshits:1979,Brailsford:1972}
\begin{align}
\frac{2\pi}{\hbar^2}|{\cal V}_{ss'}(\bfk,\bfki)|^2\d(\w_s(k) -\w_{s'}(k') -\W_q)
\end{align}
by Fermi's Golden Rule. Taking account of the energy 
$\hbar[\w_s(k) -\w_{s'}(k') ]=\hbar\W_q$ transferred for every such transition, and the initial and final state distributions of phonons, we obtain
\be
D = \frac{\pi}{\hbar}\sum_{ss'}\! \intk\!\!\intki\,\W_q\, \big|{\cal V}_{ss'}(\bfk,\bfki)\big|^2
\left[n\big(\w_{s'}(k')\big)-n\big(\w_s(k)\big)\right] \d\big(\w_s(k) -\w_{s'}(k') -\W_q\big)
\label{eq:dissipation-alshits1979}
\ee
for the energy dissipated per unit time per unit length of the moving dislocation by its transferring energy irreversibly to the phonons in the solid,
after dividing by $2$ to symmetrize over the labels on the initial and final phonon states. Here
\be
n(\w) = \frac{1}{\exp(\hbar\w/k_BT)-1}
\ee
is the finite temperature Bose-Einstein distribution function for the phonons.


Since $\W_q$ is already linear in the dislocation velocity, taking the limit $\lim\limits_{v\to 0} {\cal V}_{ss'}(\bfk, \bfki)$ gives the lowest-order 
approximation for the drag coefficient at small velocities~\cite{Alshits:1979,Alshits:1992}. The general expression (\ref{eq:dissipation-alshits1979}) is 
valid for finite velocities, provided we use the $v$-dependent vertex ${\cal V}_{ss'}(\bfk, \bfki)$ of (\ref{eq:defGamma}), with the 
dislocation strains of \eqref{eq:strains-nocutoffs}, \eqref{eq:strains-screws-nocutoffs}. Since the general expression is quite formidable,
and involves interactions with both two transverse and two longitudinal phonons, as well as mixed terms of one transverse and
one longitudinal phonon, we focus on providing numerical results for some special illustrative cases. 

Some simplification occurs in the case that only the transverse phonon contributions to the drag coefficient are considered,
due to the fact that in the square of ${\cal V}_{ss'}(\bfk, \bfki)$ and sum over the two transverse modes, the relation 
\be
\sum_{s =1,2} \bfe_i(\bk,s)\,\bfe_j(\bk,s)= \d_{ij} - \frac{\bfk_i\bfk_j}{\bfk^2}
\ee
may be used and this projector transverse to the phonon momentum $\bfk$ leads to the vanishing of a number of terms in the
sums over the third order tensor structures in (\ref{V3t}). As a result the drag coefficient due to transverse phonon wind is
independent of $3a=\ml-\mm+\frac{\mn}2$, and most terms proportional to $\tilde b=\mm-\frac{\mn}{2}$ and $\l$ drop out as well.
The drag coefficient for screw dislocations interacting with purely transverse phonons depends only on the two elastic constants 
$\m$, $\mn$, whereas dislocation drag for edge dislocations interacting with purely transverse phonons depends on the four 
elastic constants $\l$, $\mu$, $\mm$, and $\mn$.

It is also possible to obtain some simplification in the high velocity $v \rightarrow \ct$ limit where the dislocation
speed approaches the speed of (transverse) sound, and the high temperature limit $\kb T \gg \hbar\ct q_{_{BZ}}/(2\pi)$, where
$q_{_{BZ}}$ is the maximum momentum in the Brillouin zone. In this limit and in the case of screw dislocations 
scattering transverse phonons, the drag coefficient has no divergence and
\begin{align}
B \propto  \frac{b^2 \kb Tq_{_{BZ}}^4 }{\ct}
\,. \label{Bestscrewtrans}
\end{align}
In contrast, for the case of edge dislocations scattering transverse phonons the drag coefficient is divergent in the limit $v\to\ct$ and
\begin{equation}
B \sim \frac{1}{\sqrt{1 - v^2 /\ct^2 }} \, .
\end{equation}

Likewise, simplifications can be made in the low velocity limit:
A fully analytic result for $B$ from transverse phonons in the simplifying limits of low velocity and high temperature is given in the appendix, \eqnref{eq:B-smallv}, up to quadratic order in $\bt\ll1$.
Further details on these calculations and both limits can be found in the internal report Ref. \cite{Blaschke:BpaperRpt}.

\section{Numerical Results for the Drag Coefficient Due to Transverse Phonons}
\label{sec:results}

In this section we present partial results of our numerical calculations of the drag coefficient in the isotropic approximation for polycrystalline metals 
whose grains are either face-centered cubic (fcc) or body-centered cubic (bcc).
The effective isotropic elastic constants that we use as input data are assembled in Table~\ref{tab:values-metals} and were taken from Refs.~\cite{Seeger:1960,Wasserbaech:1990,Lubarda:1997}. The unit cell volumes $\Vc=a^3$ (resp. lattice parameters) were taken from Ref.~\cite[Sec. 12]{Haynes:2017}.

\begin{table}[h!t!b]
{\renewcommand{\arraystretch}{1.1}
\centering
 \begin{tabular}{c|c|c|c|c}
          & Al\,(fcc) & Cu\,(fcc) & Fe\,(bcc) & Nb\,(bcc) \\\hline
 $a$[\r{A}] & 4.05 & 3.61 & 2.87 & 3.30 \\
 $q_{_{BZ}}$[\r{A}$^{-1}$] & 0.96 & 1.08 & 1.36 & 1.18 \\
 $\rh$[kg/m$^3$]     & 2700 & 8960 & 7870 & 8570 \\\hline
 $\lambda$[GPa] & 58.1 & 105.5 & 115.5 & 144.5 \\
 $\mu$[GPa] & 26.1 & 48.3 & 81.6 & 37.5 \\\hline
 $\ml$[GPa] & $-143\pm13$ & $-160\pm70$ & $-170\pm40$ & $-610\pm80$ \\
 $\mm$[GPa] & $-297\pm6$ & $-620\pm10$ & $-770\pm10$ & $-220\pm30$ \\
 $\mn$[GPa] & $-345\pm4$ & $-1590\pm20$ & $-1520\pm10$ & $-300\pm20$ 
 \end{tabular}
 \caption{We list the experimental values used in the computation of the drag coefficient. The lattice parameters $a$ and densities $\rh$ were taken from Ref.~\cite[Sec. 12]{Haynes:2017}. The Lam\'e constants were taken from Refs.~\cite{Lubarda:1997}, \cite[p.~10]{Hertzberg:2012}.  The Murnaghan constants for Cu and Fe were taken from~\cite{Seeger:1960}, those for Al were taken from Reddy 1976 as reported by Wasserb{\"a}ch in Ref.~\cite{Wasserbaech:1990}, and those for Nb were taken from~\cite{Graham:1968}.
 Uncertainties (as given in those references) are listed as well.
 For the unit cell volume we use $\Vc =a^3$, the radius of the Brillouin zone follows from $q_{_{BZ}}=\sqrt[3]{\frac{3}{4\pi}\frac{(2\pi)^3}{\Vc}}=\sqrt[3]{6\pi^2}/a$, and $b=a/\sqrt{2}$ for fcc metals and $b=a\sqrt{3}/2$ for bcc metals (see Refs.~\cite{Frank:1958} and~\cite[Sec. 9]{Hirth:1982} for a discussion of Burgers vectors in various crystals).}
 \label{tab:values-metals}}
\end{table}

\begin{figure}[h!t!b]
 \centering
 \vspace*{-0.2cm}
 \includegraphics[width=0.84\textwidth]{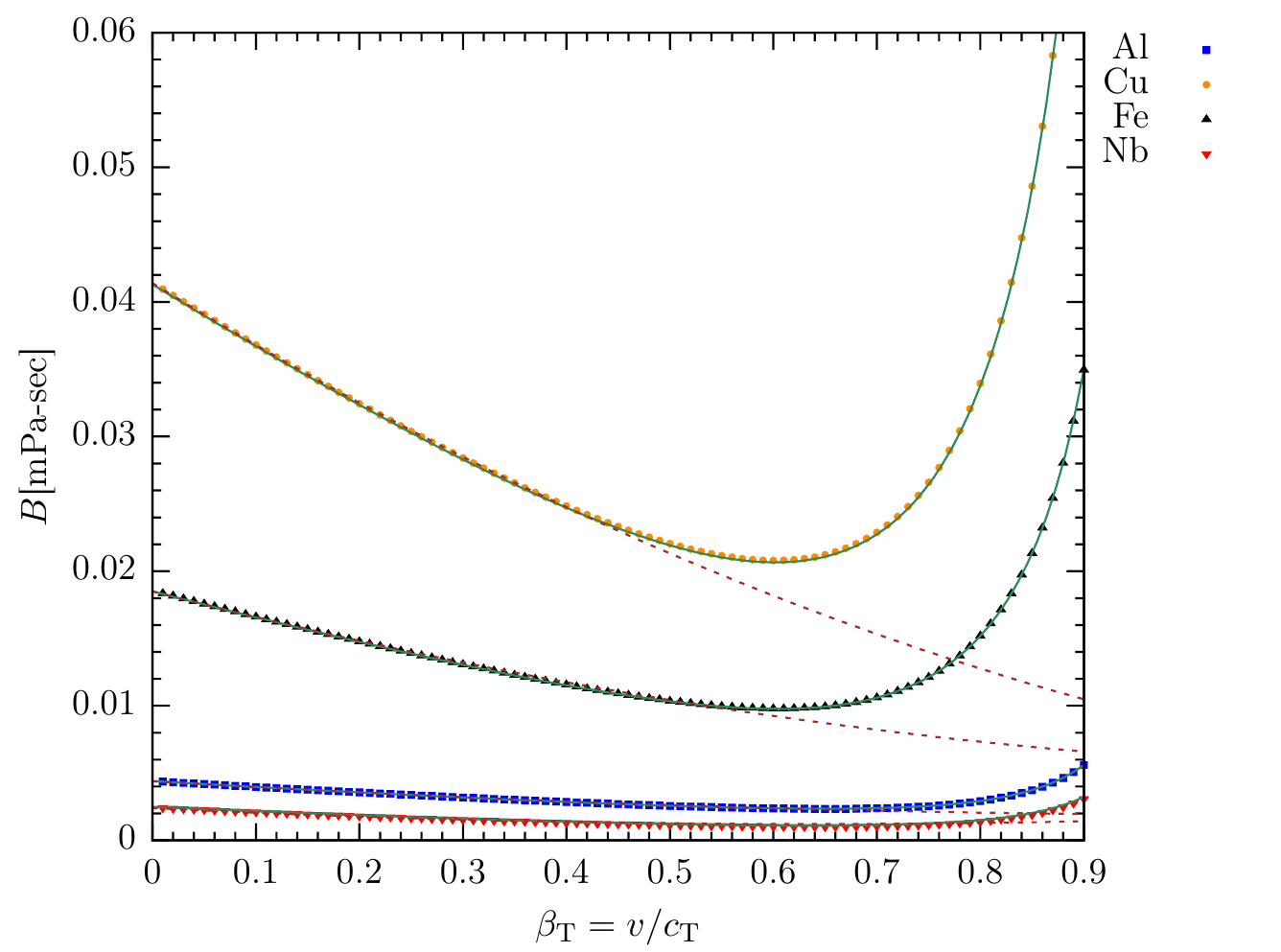}
 \vspace*{-0.2cm}
\caption{%
The room-temperature drag coefficient from phonon wind for edge dislocations in Al, Cu, Fe and Nb in the Debye approximation with zero dislocation core size (i.e. no cutoff) and with isotropic second- and third-order elastic constants from Table~\ref{tab:values-metals}.
These numerical results (Mathematica) only take into account the interaction with transverse phonons.
The numerical results (points) are overlain with least-squares-fitted curves (solid lines) of the form given in \eqnref{eq:fittedcurves} with coefficients from Table~\ref{tab:fitting-parameters}.
The dashed lines show the small velocity (large temperature) approximation of \eqnref{eq:B-smallv}, \eqref{eq:B-edge-smallv}.}
 \label{fig:dragvarious-edge}
\end{figure}

\begin{figure}[h!t!b]
 \centering
 \vspace*{-0.2cm}
 \includegraphics[width=0.84\textwidth]{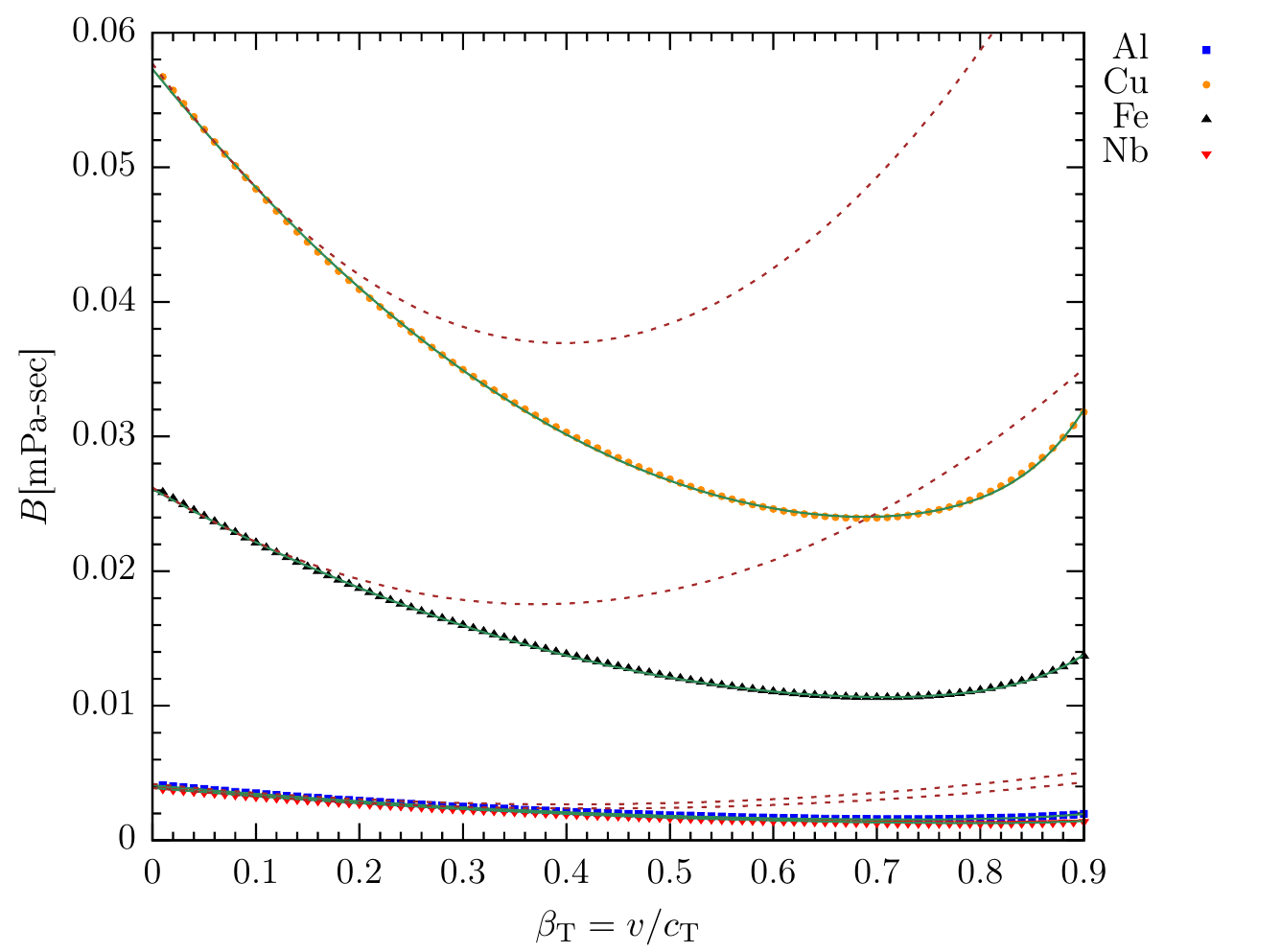}
 \vspace*{-0.2cm}
\caption{%
The room-temperature drag coefficient from phonon wind for screw dislocations in Al, Cu, Fe and Nb in the Debye approximation with zero dislocation core size (i.e. no cutoff) and with isotropic second- and third-order elastic constants from Table~\ref{tab:values-metals}.
As for edge dislocations, these numerical results only account for the interaction with transverse phonons.
Again, the solid lines are the least-squares-fitted curves of the form given in \eqnref{eq:fittedcurves} with coefficients from Table~\ref{tab:fitting-parameters}, and the numerical results are individual points.
The dashed lines show the small velocity (large temperature) approximation of \eqnref{eq:B-smallv}, \eqref{eq:B-screw-smallv}.}
 \label{fig:dragvarious-screw}
\end{figure}


The predominant sources of uncertainty are the uncertainties in the elastic constants and the size and shape to take for the dislocation core cutoff (which can significantly affect the magnitude and shape of the $B(v)$ curve unless it is much smaller than a Burgers vector).

The drag coefficients for edge and screw dislocations are plotted in Figures~\ref{fig:dragvarious-edge} and \ref{fig:dragvarious-screw}.
The Mathematica numerical integrations included only the interaction with transverse phonons.
The curves start at $\bt =0.01$, i.e. around the speed where the dislocation velocity typically becomes linearly dependent on the applied stress~\cite{Nadgornyi:1988}, hence the motion can be described as ``viscous''.
The initial small to moderate decrease in the magnitude of $B$ is described by the next to leading order term in the low velocity expansion which is linear in velocity and has a negative coefficient; see \eqnref{eq:B-smallv}.
Mathematically, the reason this coefficient is negative is that the integration range over the phonon spectrum shrinks with increasing velocity as a consequence of the energy conserving delta function in \eqref{eq:dissipation-alshits1979} in conjunction with the condition that all phonon wave vectors lie within the first Brillouin zone.
Physically, one can imagine that as the dislocation picks up speed, it is less affected by those phonons that do not hit it head on.
The magnitude of this effect depends on the material constants involved and as such seems to be more pronounced in Cu and Fe than it is in Al and Nb; see Figures~\ref{fig:dragvarious-edge} and \ref{fig:dragvarious-screw}.

At high velocities (around 70\% transverse sound speed) the drag coefficient for edge dislocations starts to grow with dislocation velocity, since the displacement gradients \eqref{edgedis} diverge at $v\to\ct$.
Hence not surprisingly, close to sound speed this growth becomes very steep, but our numerical evaluation also becomes less accurate in this limit.
For this reason the results are displayed only up to 90\% transverse sound speed.
As pointed out above, the $v$-$B$ curve is sensitive to the core cutoff and elastic constants.
In particular, differences in the elastic constants are responsible for the significantly smaller variation in $B$ for Al and Nb than for Cu and Fe for velocities up to $0.9\ct$.
Similar differences in the velocity dependence of $B$ are seen in experiments:
While experiments carried out in the velocity regime $0.1<\bt <0.5$ for LiF crystals~\cite{Cotner:1964} already show a growing drag coefficient, other experiments done up to $\bt \sim 0.7$ for NaCl crystals~\cite{Darinskaya:1982} exhibit a linear stress-velocity dependence even in this regime.

All curves shown in Figures~\ref{fig:dragvarious-edge} and~\ref{fig:dragvarious-screw} can be fit quite accurately using functions of the form
\begin{align}
 B_\txt{e}(\bt)&\approx C^\txt{e}_0 +C^\txt{e}_1\bt +C^\txt{e}_2\bt^2 +C^\txt{e}_3\log\!\left(1-\bt^2\right) +C^\txt{e}_4\left(\frac{1}{\sqrt{1-\bt^2}}-1\right)
 \,,\nn\\*
 B_\txt{s}(\bt)&\approx C^\txt{s}_0 +C^\txt{s}_1\bt +C^\txt{s}_2\bt^2 +C^\txt{s}_3\bt^4 +C^\txt{s}_4\bt^{16}
 \,, \label{eq:fittedcurves}
\end{align}
for edge ($B_\txt{edge}$) and screw ($B_\txt{screw}$) dislocations.
The forms of these functions were motivated by the expected asymptotic behavior as $\bt\to1$ in accordance with the estimates of the
previous section. Explicit values for the coefficients $C_\txt{0-3}^\txt{e/s}$ for edge/screw dislocations for Al, Cu, Fe, and Nb were computed using least squares fits; the values are summarized in Table 2.
The resulting curves are overlain with the numerical data in Figures~\ref{fig:dragvarious-edge} and~\ref{fig:dragvarious-screw}.

\begin{table}[h!t!b]
{\renewcommand{\arraystretch}{1.1}
\centering
 \begin{tabular}{c|c|c|c|c}
          & Al\,(fcc) & Cu\,(fcc) & Fe\,(bcc) & Nb\,(bcc) \\\hline
 $C_0^{\text{e}}$ & 0.0044 & 0.0414 & 0.0186 & 0.0024 \\
 $C_1^{\text{e}}$ & -0.0044 & -0.0470 & -0.0197 & -0.0035 \\
 $C_2^{\text{e}}$ & 0.0025 & 0.0233 & 0.0110 & 0.0030 \\
 $C_3^{\text{e}}$ & 0.0070 & 0.1032 & 0.0458 & 0.0052 \\
 $C_4^{\text{e}}$ & 0.0114 & 0.1809 & 0.0783 & 0.0078 \\
 \hline
 $C_0^{\text{s}}$ & 0.0041 & 0.0573 & 0.0261 & 0.0039 \\
 $C_1^{\text{s}}$ & -0.0069 & -0.0946 & -0.0425 & -0.0065 \\
 $C_2^{\text{s}}$ & 0.0047 & 0.0667 & 0.0289 & 0.0043 \\
 $C_3^{\text{s}}$ & -0.0001 & 0.0008 & 0.0004 & -0.0005 \\
 $C_4^{\text{s}}$ & 0.0018 & 0.0285 & 0.0122 & 0.0013 \\
 \hline\hline
 $B_{\text{e}}$ & 0.0044 & 0.0409 & 0.0184 & 0.0024 \\
 $B_{\text{s}}$ & 0.0041 & 0.0567 & 0.0259 & 0.0038 \\
 \end{tabular}
\caption{Fitting parameters $C^{\txt{e}}_m$/$C^{\txt{s}}_m$ for edge/screw dislocations in units of mPa-sec.
In the last two lines we report (again in units of mPa-sec) the numerically computed results for the drag coefficient at $\bt=0.01$.}
\label{tab:fitting-parameters}
}
\end{table}

\begin{figure}[ht]
 \centering
 \includegraphics[width=0.55\textwidth]{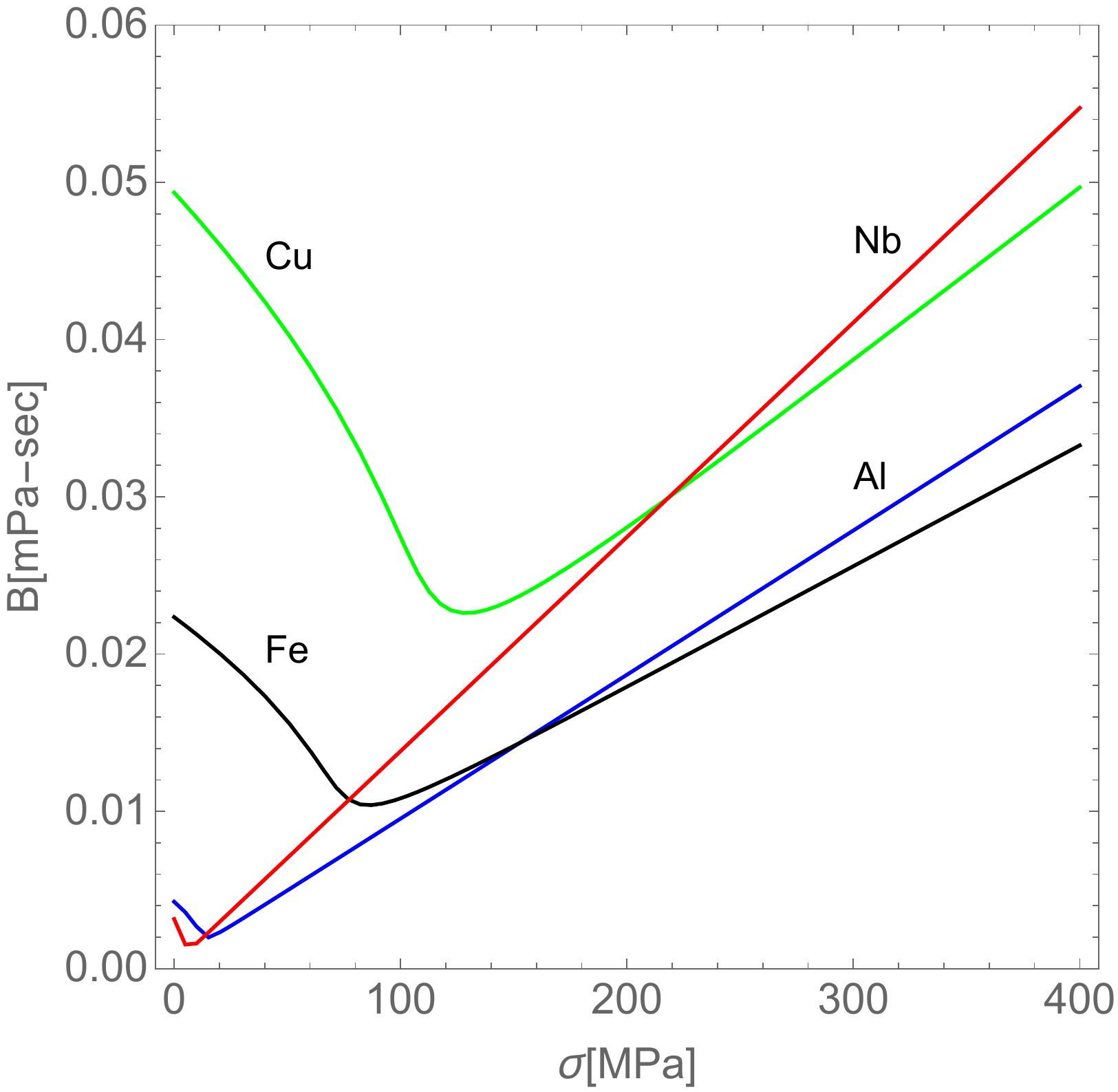}
 \caption{We show $B=(B_\txt{e}+B_\txt{s})/2$ as a function of effective shear stress $\sigma$ determined from \eqref{eq:fittedcurves}.}
 \label{fig:Bofsigma}
\end{figure}

So far, we considered the dislocation drag coefficient as a function of velocity.
Since the dislocation velocity is related to the effective (local) shear stress according to $b\,\sigma = vB(v)$, we can numerically solve for $v(\sigma)$ which gives $B$ as a function of $\sigma$:
$B(\sigma)\coleq B(v(\sigma))$.
Figure \ref{fig:Bofsigma} shows the mean value $B=(B_\txt{e}+B_\txt{s})/2$ as a function of the effective shear stress.
In the high stress regime, where $B$ becomes linear in stress, $v(\sigma)$ is already close to $\ct$ and edge dislocations dominate; see \eqref{eq:fittedcurves}.


Comparing our results in the low-velocity regime without any core cutoff (cf. Table~\ref{tab:values-metals}) to experiments and MD simulations, we find that our drag coefficient for
\begin{itemize}
\itemsep=0pt
 \item Al is below the range of experimental values of $\sim0.02\,$mPa-sec in~\cite{Gorman:1969} and $\sim0.06\,$mPa-sec in~\cite{Parameswaran:1972}, and also below the MD simulation values of $\sim0.01\,$mPa-sec for edge and $\sim0.02\,$mPa-sec for screw dislocations in~\cite{Olmsted:2005},
 \item Cu is well within the range of experimental values of $\sim0.0079\,$mPa-sec in~\cite{Suzuki:1964}, $\sim0.02\,$mPa-sec in~\cite{Jassby:1973,Zaretsky:2013},
 $\sim0.065\,$mPa-sec (for both edge and screw dislocations) in~\cite{Stern:1962},
 $\sim0.07\,$mPa-sec in~\cite{Greenman:1967},
 and $\sim0.08\,$mPa-sec in~\cite{Alers:1961},
 \item Fe is lower than the experimental values of $\sim0.34\,$mPa-sec for edge and $\sim0.661\,$mPa-sec for screw dislocations reported in~\cite{Urabe:1975},
 as well as the result $\sim0.26\,$mPa-sec of MD simulations for screw dislocations reported in~\cite{Gilbert:2011}.
\end{itemize}
We did not find experimental data for Nb.
All results above were computed for zero pressure and room temperature, i.e. $300\,$K, because the elastic constants we used were measured at this temperature; see Table~\ref{tab:values-metals}.
We note also, that the experiments mentioned above typically were unable to distinguish between edge and screw dislocations.
Furthermore, dislocations moving at speeds comparable to $\ct$ cannot be measured directly in experiments.

In comparing the high velocity behavior with MD simulations we note that the limiting velocity changes from $\ct$ to a slip system and dislocation character dependent shear wave speed when anisotropic effects are taken into account~\cite{Blaschke:2017lten,Blaschke:2019fits}.
Furthermore, the degree of divergence is enhanced for steady state dislocations in this case~\cite{Blaschke:2018anis,Blaschke:2019fits}.
On the other hand, the inclusion of a constant acceleration term into the isotropic dislocation field has been shown to reduce the degree of divergence~\cite{Markenscoff:2008,Markenscoff:2009,Huang:2009}.
Therefore, our present results cannot be expected to match MD simulations in this regime.
Nonetheless, we note that Refs. \cite{Olmsted:2005,Oren:2017} show that screw dislocations in fcc aluminum and copper can accelerate to transonic speeds, consistent with our present asymptotic form \eqref{Bestscrewtrans}.
Edge dislocations, on the other hand, seem to approach a limiting velocity up to a critical driving stress above which they too become transonic in MD simulations.
The latter behavior cannot be described by our present model.

\section{Summary}
\label{sec:discussion}

In this paper we have studied the velocity dependence of the dislocation drag coefficient for dislocation velocities $v$ in the range $0.01\ct <v<\ct $ where $\ct $ is the transverse sound speed.  In this regime the dominant contribution to dislocation drag is the dissipative interaction with phonons, i.e. the phonon wind.
Although the currently employed model breaks down at $v=\ct $, we were able to make predictions for the velocity dependence of the drag coefficient $B(v)$ at dislocation speeds below this critical value.
Our main results are shown in Figures~\ref{fig:dragvarious-edge} and \ref{fig:dragvarious-screw} for edge and screw dislocations respectively, for four different metals Al, Cu, Fe, Nb, chosen for their simple lattice structure and available data for their isotropic third-order elastic constants at room temperature.
We computed $B(v)$ in the range $v/\ct \in[0.01,0.9]$ and can represent all results by simple fitting functions of the form \eqnref{eq:fittedcurves} with five fitting parameters.
We have compared our results to experimental values and MD simulation results where these are available, i.e. in the low-velocity regime, i.e. $v/\ct \sim 0.01$.
We found good agreement for copper, while our results for aluminum and iron are lower in the low-velocity regime.
Additional experimental data on third-order elastic constants is necessary to improve our predictions and to compute the drag coefficient in the isotropic limit for other materials.

\subsection*{Acknowledgements}

We thank Darby J. Luscher and Benjamin A. Szajewski for enlightening discussions
and the anonymous referees for their valuable comments.
This work was performed under the auspices of the U.S. Department of Energy under contract 89233218CNA000001.
In particular, the authors are grateful for the support of the Advanced Simulation and Computing, Physics and Engineering Models Program.


\providecommand{\accepted}[1]{accepted for publication in \textit{#1}}
\providecommand{\href}[2]{#2}\begingroup\begin{thebibliography}{10}
\small\itemsep=3pt
\tolerance 1414
\hbadness 1414
\emergencystretch 1.5em
\hfuzz 0.3pt
\widowpenalty=10000
\vfuzz \hfuzz
\raggedbottom

\bibitem{Hunter:2015}
A.~Hunter and D.~L. Preston, ``Analytic model of the remobilization of pinned
  glide dislocations from quasi-static to high strain rates'',
  \href{https://dx.doi.org/10.1016/j.ijplas.2015.01.008}{\emph{Int. J. Plast.}
  \textbf{70} (2015) 1--29}.

\bibitem{Nadgornyi:1988}
E.~M. Nadgornyi, ``Dislocation dynamics and mechanical properties of
  crystals'',
  \href{https://dx.doi.org/10.1016/0079-6425(88)90005-9}{\emph{Prog. Mater.
  Sci.} \textbf{31} (1988) 1--530}.

\bibitem{Clifton:1971}
R.~J. Clifton, ``On the analysis of elastic/visco-plastic waves of finite
  uniaxial strain'', in \emph{Shock Waves and the Mechanical Properties of
  Solids}, J.~J. Burke and V.~Weiss, eds., vol.~17 of \emph{Sagamore Army
  Materials Research Conference Proceedings}, pp.~73--116, (Syracuse, N.Y.:
  Syracuse University Press, 1971).

\bibitem{Krasnikov:2010}
V.~S. Krasnikov, A.~{\relax Yu}. Kuksin, A.~E. Mayer, and A.~V. Yanilkin,
  ``Plastic deformation under high-rate loading: {The} multiscale approach'',
  \href{https://dx.doi.org/10.1134/S1063783410070115}{\emph{Phys. Solid State}
  \textbf{52} (2010) 1386--1396}.

\bibitem{Barton:2011}
N.~R. Barton, J.~V. Bernier, R.~Becker, \emph{et~al.}, ``A multiscale strength
  model for extreme loading conditions'',
  \href{https://dx.doi.org/10.1063/1.3553718}{\emph{J. Appl. Phys.}
  \textbf{109} (2011) 073501}.

\bibitem{Luscher:2016}
D.~J. Luscher, J.~R. Mayeur, H.~M. Mourad, A.~Hunter, and M.~A. Kenamond,
  ``Coupling continuum dislocation transport with crystal plasticity for
  application to shock loading conditions'',
  \href{https://dx.doi.org/10.1016/j.ijplas.2015.07.007}{\emph{Int. J. Plast.}
  \textbf{76} (2016) 111--129}.

\bibitem{Austin:2018}
R.~A. Austin, ``Elastic precursor wave decay in shock-compressed aluminum over
  a wide range of temperature'',
  \href{https://dx.doi.org/10.1063/1.5008280}{\emph{J. Appl. Phys.}
  \textbf{123} (2018) 035103}.

\bibitem{Alshits:1992}
V.~I. Alshits,
  \href{https://dx.doi.org/10.1016/B978-0-444-88773-3.50018-2}{``The
  phonon-dislocation interaction and its role in dislocation dragging and
  thermal resistivity'',} in \emph{Elastic Strain Fields and Dislocation
  Mobility}, V.~L. Indenbom and J.~Lothe, eds., vol.~31 of \emph{Modern
  Problems in Condensed Matter Sciences}, pp.~625--697, (Elsevier, 1992).

\bibitem{Lothe:1960}
J.~Lothe, ``Aspects of the theories of dislocation mobility and internal
  friction'', \href{https://dx.doi.org/10.1103/PhysRev.117.704}{\emph{Phys.
  Rev.} \textbf{117} (1960) 704--708}.

\bibitem{Lothe:1962}
J.~Lothe, ``Theory of dislocation mobility in pure slip'',
  \href{https://dx.doi.org/10.1063/1.1728907}{\emph{J. Appl. Phys.} \textbf{33}
  (1962) 2116--2125}.

\bibitem{Blaschke:BpaperRpt}
D.~N. Blaschke, E.~Mottola, and D.~L. Preston,
  \href{https://dx.doi.org/10.2172/1434423}{``On the velocity dependence of the
  dislocation drag coefficient from phonon wind'',} Tech. Rep. LA-UR-16-24559,
  Los Alamos Natl. Lab., 2018.

\bibitem{Rosakis:2001}
P.~Rosakis, ``Supersonic dislocation kinetics from an augmented {Peierls}
  model'', \href{https://dx.doi.org/10.1103/PhysRevLett.86.95}{\emph{Phys. Rev.
  Lett.} \textbf{86} (2001) 95--98}.

\bibitem{Marian:2006}
J.~Marian and A.~Caro, ``Moving dislocations in disordered alloys: {Connecting}
  continuum and discrete models with atomistic simulations'',
  \href{https://dx.doi.org/10.1103/PhysRevB.74.024113}{\emph{Phys. Rev.}
  \textbf{B74} (2006) 024113}.

\bibitem{Pellegrini:2010}
Y.-P. Pellegrini, ``Dynamic {Peierls-Nabarro} equations for elastically
  isotropic crystals'',
  \href{https://dx.doi.org/10.1103/PhysRevB.81.024101}{\emph{Phys. Rev.}
  \textbf{B81} (2010) 024101},
  \href{https://arxiv.org/abs/0908.2371}{\texttt{arXiv:0908.2371
  [cond-mat.mtrl-sci]}}.

\bibitem{Ruestes:2015}
C.~J. Ruestes, E.~M. Bringa, R.~E. Rudd, B.~A. Remington, T.~P. Remington, and
  M.~A. Meyers, ``Probing the character of ultra-fast dislocations'',
  \href{https://dx.doi.org/10.1038/srep16892}{\emph{Sci. Rep.} \textbf{5}
  (2015) 16892}.

\bibitem{Oren:2017}
E.~Oren, E.~Yahel, and G.~Makov, ``Dislocation kinematics: a molecular dynamics
  study in {Cu}'',
  \href{https://dx.doi.org/10.1088/1361-651X/aa52a7}{\emph{Mod. Simul. Mater.
  Sci. Eng.} \textbf{25} (2017) 025002}.

\bibitem{Nosenko:2007}
V.~Nosenko, S.~Zhdanov, and G.~Morfill, ``Supersonic dislocations observed in a
  plasma crystal'',
  \href{https://dx.doi.org/10.1103/PhysRevLett.99.025002}{\emph{Phys. Rev.
  Lett.} \textbf{99} (2007) 025002},
  \href{https://arxiv.org/abs/0709.1782}{\texttt{arXiv:0709.1782
  [cond-mat.soft]}}.

\bibitem{Blaschke:2017lten}
D.~N. Blaschke and B.~A. Szajewski, ``Line tension of a dislocation moving
  through an anisotropic crystal'',
  \href{https://dx.doi.org/10.1080/14786435.2018.1489152}{\emph{Phil. Mag.}
  \textbf{98} (2018) 2397--2424},
  \href{https://arxiv.org/abs/1711.10555}{\texttt{arXiv:1711.10555
  [cond-mat.mtrl-sci]}}.

\bibitem{Blaschke:2018anis}
D.~N. Blaschke, ``Velocity dependent dislocation drag from phonon wind and
  crystal geometry'',
  \href{https://dx.doi.org/10.1016/j.jpcs.2018.08.032}{\emph{J. Phys. Chem.
  Solids} \textbf{124} (2019) 24--35},
  \href{https://arxiv.org/abs/1804.01586}{\texttt{arXiv:1804.01586
  [cond-mat.mtrl-sci]}}.

\bibitem{Blaschke:2019fits}
D.~N. Blaschke, ``Properties of dislocation drag from phonon wind at ambient
  conditions'', \href{https://dx.doi.org/10.3390/ma12060948}{\emph{Materials}
  \textbf{12} (2019) 948},
  \href{https://arxiv.org/abs/1902.02451}{\texttt{arXiv:1902.02451
  [cond-mat.mtrl-sci]}}.

\bibitem{Wallace:1972}
D.~C. Wallace, \emph{Thermodynamics of Crystals}, (New York: J. Wiley \& Sons
  Inc., 1972).

\bibitem{Eshelby:1949}
J.~D. Eshelby, ``Uniformly moving dislocations'',
  \href{https://dx.doi.org/10.1088/0370-1298/62/5/307}{\emph{Proc. Phys. Soc.}
  \textbf{A62} (1949) 307}.

\bibitem{Weertman:1980}
J.~Weertman and J.~R. Weertman, ``Moving dislocations'', in \emph{Moving
  Dislocations}, F.~R.~N. Nabarro, ed., vol.~3 of \emph{Dislocations in
  Solids}, pp.~1--59, (Amsterdam: North Holland Pub. Co., 1980).

\bibitem{Landau:1986}
L.~Landau and E.~Lifshitz, \emph{Theory of Elasticity}, third~ed., vol.~7 of
  \emph{Course of Theoretical Physics}, (Butterworth-Heinemann, 1986).

\bibitem{Kosevich:2005}
A.~M. Kosevich, \href{https://dx.doi.org/10.1002/352760667X}{\emph{The Crystal
  Lattice: Phonons, Solitons, Dislocations, Superlattices}}, 2nd~ed.,
  (Wiley-VCH, 2005).

\bibitem{Teodosiu:1982}
C.~Teodosiu, \href{https://dx.doi.org/10.1007/978-3-662-11634-0}{\emph{Elastic
  Models of Crystal Defects}}, (Springer-Verlag, 1982).

\bibitem{Hirth:1982}
J.~P. Hirth and J.~Lothe, \emph{Theory of Dislocations}, second~ed., (New York:
  Wiley, 1982).

\bibitem{Murnaghan:1937}
F.~D. Murnaghan, ``Finite deformations of an elastic solid'',
  \href{https://dx.doi.org/10.2307/2371405}{\emph{Am. J. Math} \textbf{59}
  (1937) 235--260}.

\bibitem{Murnaghan:1951}
F.~D. Murnaghan, \emph{Finite Deformations of an Elastic Solid}, (New York:
  Wiley, 1951).

\bibitem{Toupin:1961}
R.~A. Toupin and B.~Bernstein, ``Sound waves in deformed perfectly elastic
  materials. {Acoustoelastic} effect'',
  \href{https://dx.doi.org/10.1121/1.1908623}{\emph{J. Acoust. Soc. Am.}
  \textbf{33} (1961) 216--225}.

\bibitem{Volkov:2015}
A.~D. Volkov, A.~I. Kokshaiskii, A.~I. Korobov, and V.~M. Prokhorov, ``Second-
  and third-order elastic coefficients in polycrystalline aluminum alloy
  {AMg6}'', \href{https://dx.doi.org/10.1134/S1063771015060147}{\emph{Acoust.
  Phys.} \textbf{61} (2015) 651--656}.

\bibitem{Lubliner:2008}
J.~Lubliner, \emph{Plasticity Theory}, {Dover}~ed., Dover Books on Engineering,
  (Dover, 2008).

\bibitem{Wallace:1970}
D.~C. Wallace,
  \href{https://dx.doi.org/10.1016/S0081-1947(08)60010-7}{``Thermoelastic
  theory of stressed crystals and higher-order elastic constants'',} in vol.~25
  of \emph{Solid State Physics}, pp.~301--404, H.~Ehrenreich, F.~Seitz, and
  D.~Turnbull, eds., (New York: Academic Press, 1970).

\bibitem{Lubarda:1997}
V.~A. Lubarda, ``New estimates of the third-order elastic constants for
  isotropic aggregates of cubic crystals'',
  \href{https://dx.doi.org/10.1016/S0022-5096(96)00113-5}{\emph{J. Mech. Phys.
  Solids} \textbf{45} (1997) 471--490}.

\bibitem{Blaschke:2017Poly}
D.~N. Blaschke, ``Averaging of elastic constants for polycrystals'',
  \href{https://dx.doi.org/10.1063/1.4993443}{\emph{J. Appl. Phys.}
  \textbf{122} (2017) 145110},
  \href{https://arxiv.org/abs/1706.07132}{\texttt{arXiv:1706.07132
  [cond-mat.mtrl-sci]}}.

\bibitem{Burgers:1939a}
J.~M. Burgers, ``Some considerations on the fields of stress connected with
  dislocations in a regular crystal lattice. {I}'',
  \href{http://www.dwc.knaw.nl/DL/publications/PU00017316.pdf}{\emph{Proc. Kon.
  Ned. Akad. v. Wet.} \textbf{42} (1939) 293--325}.

\bibitem{Weertman:1961}
J.~Weertman, ``High velocity dislocations'', in \emph{Response of Metals to
  High Velocity Deformation}, P.~G. Shewmon and V.~F. Zackay, eds., vol.~9 of
  \emph{Metallurgical Society Conferences}, pp.~205--247, (New York:
  Interscience Publishers, 1961).
\newblock Proc. of a technical conference, Estes Park, Colorado, 1960.

\bibitem{Alshits:1979}
V.~I. Al'shits, M.~D. Mitlianskij, and R.~K. Kotowski, ``The phonon wind as a
  non-linear mechanism of dislocation dragging'', \emph{Arch. Mech.}
  \textbf{31} (1979) 91--105.

\bibitem{Alshits:1969a}
V.~I. Al'shits, ``Raman scattering of phonons as a cause of dislocation
  damping'', \emph{Sov. Phys. Solid State} \textbf{11} (1969) 1081--1087,
  [\textit{Fiz. Tverd. Tela} \textbf{11} (1969) 1336--1344].

\bibitem{Alshits:1969b}
V.~I. Al'shits, ``{`Phonon wind'} and dislocation damping'', \emph{Sov. Phys.
  Solid State} \textbf{11} (1970) 1947--1948, [\textit{Fiz. Tverd. Tela}
  \textbf{11} (1969) 2405--2407].

\bibitem{Alshits:1973}
V.~I. Al'shitz and A.~G. Mal'shukov, ``Phonon component of dynamic dragging of
  dislocations'',
  \href{http://www.jetp.ac.ru/cgi-bin/dn/e_036_05_0978.pdf}{\emph{Sov. Phys.
  JETP} \textbf{36} (1973) 978--982}, [\textit{Zh. Eksp. Teor. Fiz.}
  \textbf{63} (1972) 1849--1857].

\bibitem{Brailsford:1972}
A.~D. Brailsford, ``Anharmonicity contributions to dislocation drag'',
  \href{https://dx.doi.org/10.1063/1.1661329}{\emph{J. Appl. Phys.} \textbf{43}
  (1972) 1380--1393}.

\bibitem{Seeger:1960}
A.~Seeger and O.~Buck, ``{Die experimentelle Ermittlung der elastischen
  Konstanten h{\"o}herer Ordnung}'',
  \href{http://zfn.mpdl.mpg.de/data/Reihe_A/15/ZNA-1960-15a-1056.pdf}{\emph{Z.
  Naturf.} \textbf{15a} (1960) 1056--1067}.

\bibitem{Wasserbaech:1990}
W.~Wasserb{\"a}ch, ``Third-order constants of a cubic quasi-isotropic solid'',
  \href{https://dx.doi.org/10.1002/pssb.2221590216}{\emph{phys. stat. sol. (b)}
  \textbf{159} (1990) 689--697}.

\bibitem{Haynes:2017}
W.~M. Haynes, \href{http://hbcponline.com}{\emph{CRC Handbook of Chemistry and
  Physics}}, 97th~ed., (CRC Press, 2017).

\bibitem{Hertzberg:2012}
R.~W. Hertzberg, R.~P. Vinci, and J.~L. Hertzberg, \emph{Deformation and
  Fracture Mechanics of Engineering Materials}, fifth~ed., (Wiley, 2012).

\bibitem{Graham:1968}
L.~J. Graham, H.~Nadler, and R.~Chang, ``Third-order elastic constants of
  single-crystal and polycrystalline columbium'',
  \href{https://dx.doi.org/10.1063/1.1656726}{\emph{J. Appl. Phys.} \textbf{39}
  (1968) 3025--3033}.

\bibitem{Frank:1958}
F.~C. Frank, ``Dislocation theory'',
  \href{https://dx.doi.org/10.1007/BF02751488}{\emph{Nuov. Cim.} \textbf{7}
  (1958) 386--413}.

\bibitem{Cotner:1964}
J.~Cotner and J.~Weertman, ``High dislocation velocities and the structures of
  slip bands in shock loaded high purity lithium fluoride'',
  \href{https://dx.doi.org/10.1039/DF9643800225}{\emph{Discuss. Faraday Soc.}
  \textbf{38} (1964) 225--232}.

\bibitem{Darinskaya:1982}
E.~V. Darinskaya, I.~P. Makarevich, {\relax Yu}.~I. Meshcheryakov, V.~A.
  Morozov, and A.~A. Urusovskaya, ``Investigation of the mobility of edge
  dislocations in {LiF} and {NaCl} crystals subjected to pulse loading with an
  electron beam'', \emph{Sov. Phys. Solid State} \textbf{24} (1982) 898,
  [\textit{Fiz. Tverd. Tela} \textbf{24} (1982) 1564].

\bibitem{Gorman:1969}
J.~A. Gorman, D.~S. Wood, and T.~{Vreeland Jr.}, ``Mobility of dislocations in
  aluminum'', \href{https://dx.doi.org/10.1063/1.1657472}{\emph{J. Appl. Phys.}
  \textbf{40} (1969) 833--841}.

\bibitem{Parameswaran:1972}
V.~R. Parameswaran, N.~Urabe, and J.~Weertman, ``Dislocation mobility in
  aluminum'', \href{https://dx.doi.org/10.1063/1.1661644}{\emph{J. Appl. Phys.}
  \textbf{43} (1972) 2982--2986}.

\bibitem{Olmsted:2005}
D.~L. Olmsted, L.~G. Hector~Jr., W.~A. Curtin, and R.~J. Clifton, ``Atomistic
  simulations of dislocation mobility in {Al, Ni} and {Al/Mg} alloys'',
  \href{https://dx.doi.org/10.1088/0965-0393/13/3/007}{\emph{Mod. Simul. Mater.
  Sci. Eng.} \textbf{13} (2005) 371},
  \href{https://arxiv.org/abs/cond-mat/0412324}{\texttt{arXiv:cond-mat/0412324}}.

\bibitem{Suzuki:1964}
T.~Suzuki, A.~Ikushima, and M.~Aoki, ``Acoustic attenuation studies of the
  frictional force on a fast moving dislocation'',
  \href{https://dx.doi.org/10.1016/0001-6160(64)90107-5}{\emph{Acta Met.}
  \textbf{12} (1964) 1231--1240}.

\bibitem{Jassby:1973}
K.~M. Jassby and T.~{Vreeland Jr.}, ``Dislocation mobility in pure copper at
  {4.2$^\circ$K}'',
  \href{https://dx.doi.org/10.1103/PhysRevB.8.3537}{\emph{Phys. Rev.}
  \textbf{B8} (1973) 3537--3541}.

\bibitem{Zaretsky:2013}
E.~B. Zaretsky and G.~I. Kanel, ``Response of copper to shock-wave loading at
  temperatures up to the melting point'',
  \href{https://dx.doi.org/10.1063/1.4819328}{\emph{J. Appl. Phys.}
  \textbf{114} (2013) 083511}.

\bibitem{Stern:1962}
R.~M. Stern and A.~V. Granato, ``Overdamped resonance of dislocations in
  copper'', \href{https://dx.doi.org/10.1016/0001-6160(62)90014-7}{\emph{Acta
  Met.} \textbf{10} (1962) 358--381}.

\bibitem{Greenman:1967}
W.~F. Greenman, T.~{Vreeland Jr.}, and D.~S. Wood, ``Dislocation mobility in
  copper'', \href{https://dx.doi.org/10.1063/1.1710178}{\emph{J. Appl. Phys.}
  \textbf{38} (1967) 3595--3603}.

\bibitem{Alers:1961}
G.~A. Alers and D.~O. Thompson, ``Dislocation contributions to the modulus and
  damping in copper at megacycle frequencies'',
  \href{https://dx.doi.org/10.1063/1.1735992}{\emph{J. Appl. Phys.} \textbf{32}
  (1961) 283--293}.

\bibitem{Urabe:1975}
N.~Urabe and J.~Weertman, ``Dislocation mobility in potassium and iron single
  crystals'',
  \href{https://dx.doi.org/10.1016/0025-5416(75)90071-3}{\emph{Mater. Sci.
  Eng.} \textbf{18} (1975) 41--49}.

\bibitem{Gilbert:2011}
M.~R. Gilbert, S.~Queyreau, and J.~Marian, ``Stress and temperature dependence
  of screw dislocation mobility in {$\alpha$-Fe} by molecular dynamics'',
  \href{https://dx.doi.org/10.1103/PhysRevB.84.174103}{\emph{Phys. Rev.}
  \textbf{B84} (2011) 174103}.

\bibitem{Markenscoff:2008}
X.~Markenscoff and S.~Huang, ``Analysis for a screw dislocation accelerating
  through the shear-wave speed barrier'',
  \href{https://dx.doi.org/10.1016/j.jmps.2008.01.005}{\emph{J. Mech. Phys.
  Solids} \textbf{56} (2008) 2225--2239}.

\bibitem{Markenscoff:2009}
X.~Markenscoff and S.~Huang, ``The energetics of dislocations accelerating and
  decelerating through the shear-wave speed barrier'',
  \href{https://dx.doi.org/10.1063/1.3072351}{\emph{Appl. Phys. Lett.}
  \textbf{94} (2009) 021906}.

\bibitem{Huang:2009}
S.~Huang and X.~Markenscoff, ``Is intersonic dislocation motion possible?
  {Singularity} analysis for an edge dislocation accelerating through the shear
  wave speed barrier'',
  \href{https://dx.doi.org/10.1007/s11340-008-9122-8}{\emph{Exp. Mech.}
  \textbf{49} (2009) 219--224}.

\end{thebibliography}\endgroup


\begingroup\endgroup
\providecommand{\href}[2]{#2}

\appendix
\section{Appendix: Small velocity and large temperature approximation}
\label{sec:appendix}

Since most applications involve temperatures around and above the Debye temperature, we employ a high temperature expansion up to next-to-leading order.
Upon expanding (up to next-to-leading order) for small dislocation velocities ($\bt \ll1$), we may compute the remaining integrals analytically,
and can compare the leading-order term to previous work such as~\cite{Alshits:1979}.
Additionally, we drop the mixed terms proportional to $\bt ^2/(\kb T)$, since they are small compared to the others for large $T$ and small $\bt $.
In this case we find for the interaction of edge and screw dislocations with transverse phonons (i.e. $s=\txt{T}$):
{\allowdisplaybreaks
\begin{align}
 B&\approx \frac{b^2 }{5 (8\mu) ^2} \Bigg(\Bigg[
\frac{\kb T \sqrt[3]{3} \left(\frac{\pi}2\right)^{\frac23} }{ \ct \Vc ^{\frac43}}-
 \frac{ \pi ^2 \ct \hbar^2}{6\kb T \Vc ^2 }\Bigg]f_0(\l,\m,\mm,\mn)
 \nn\\*
  &\quad\qquad
 - \frac{\bt}{\pi}\Bigg[\frac{\kb T\sqrt[3]{2}\left(\tfrac{\pi}{3}\right)^{\frac23}}{\ct\Vc^{\frac43}}-\frac{\pi^2\ct\hbar^2}{6\kb T \Vc^2}\Bigg]f_1(\l,\m,\mm,\mn)
 + \bt ^2\frac{\kb T \left(\frac{9\pi}{16}\right)^{\!\frac23} }{ 7\ct \Vc ^{\frac43}} f_2(\l,\m,\mm,\mn)
  \Bigg)
   \,, \label{eq:B-smallv}
\end{align}
where the coefficients $f_{0,1,2}$ depend only on the elastic constants $\l$, $\mu$, $\mm$, $\mn$ (not $\ml$), and their explicit form depends on whether we consider edge or screw dislocations.
In particular, for edge dislocations the coefficients read
\begin{align}
 f_0^\txt{e}(\l,\m,\mm,\mn)&=\Big[516 (2\mu)^4\!+165 \l ^2 \mn^2+16\mu^2 \big(370 \l ^2+151 \mm^2+274 \l  \mm-44 \mm \mn+15 \mn^2+119 \l  \mn\big) \nn\\*
 &\quad +96 \mu ^3 (133 \l +64 \mm+4 \mn)+2 \l  \mu  \mn (764 \l +76 \mm+141 \mn)\Big]/\big[84(\l +2 \mu )^2\big]
 \,, \nn\\
 f_1^\txt{e}(\l,\m,\mm,\mn)&= \Big[2 \mu ^2 \left(798 \lambda ^2+6 (8\mm)^2+8 \mm (42 \lambda -19 \mn)+51 \mn^2+296 \lambda 
   \mn\right) + 183 (2\mu) ^4 +\frac{111
   \lambda ^2 \mn^2}{2} \nn\\*
 &\quad +16 \mu ^3 (235 \lambda +92 \mm+11 \mn)+2 \lambda  \mu  \mn (206 \lambda -36 \mm+61 \mn)\Big]/\big[5(\l +2 \mu )^2\big]
 \,, \nn\\
 f_2^\txt{e}(\l,\m,\mm,\mn)&=\Big[4 \mu ^3 \left(6933 (4\l) ^2+2952 (4\mm)^2+520 \l  (102 \mm-7 \mn)-22076 \mm \mn+3977 \mn^2\right) \nn\\*
 &\quad +2 \l  \mu ^2 \left(1969 (4 \l)^2+8486 (2\mm)^2+4\mm(1676 \l  -8673 \mn)+9951 \mn^2+6708 \l \mn\right) \nn\\*
 &\quad +4 \l ^2 \mu  \mn (2661 \mn-38 \l -4540 \mm)+96 \mu ^4 (9021 \l +3260 \mm-259 \mn) \nn\\*
 &\quad +18814 (2\mu)^5+ 1411 \l ^3 \mn^2\Big]/\big[198(\l +2 \mu )^3\big]
 \,,\label{eq:B-edge-smallv}
\end{align}
whereas for screw dislocations they compute to
\begin{align}
 f_0^\txt{s}(\l,\m,\mm,\mn)&=\frac{1}{7} \left(33 (2\mu)^2 + 62 \mu \mn + \frac{131}{12} \mn^2\right)
 \,, \nn\\
 f_1^\txt{s}(\l,\m,\mm,\mn)&= 41 (2\mu)^2 + 76 \mu \mn + \frac{27}{2} \mn^2
 \,, \nn\\
 f_2^\txt{s}(\l,\m,\mm,\mn)&=\frac{1}{11} \bigg(\frac{3635 (2\mu)^2}{3} + 2202 \mu \mn + \frac{1457 \mn^2}{4}\bigg)
   \,. \label{eq:B-screw-smallv}
\end{align}
The simpler structure in the screw case is partly due to the deformation field depending only on $\ct $, whereas the deformation
field for edge dislocations also includes terms depending on $\cl $; see \eqref{eq:strains-nocutoffs}, \eqref{eq:strains-screws-nocutoffs}.
In fact, in  order to arrive at \eqref{eq:B-edge-smallv} we used the relation $\cl =\ct \sqrt{\frac{\l+2\m}{\m}}$.
Additionally, the coefficients in the screw case depend only on $\m$ and $\mn$.
Notice that the first term in \eqref{eq:B-smallv} qualitatively agrees with
Ref.~\cite{Alshits:1979}, albeit differing in some numerical coefficients within $f_0^{\txt{e,s}}$
\footnote{Note that here we have removed the cutoff for the dislocation core and expanded for large temperature.
Furthermore, in order to compare the two expressions we note that the wave vector cutoff denoted in~\cite{Alshits:1973} by $k_\txt{D} $ is related to the unit cell volume,
$k_\txt{D} \propto \Vc ^{-1/3}$.
}.
This discrepancy can be traced back to the tensor of third order elastic constants used in that paper which seems to be incorrect. 
}

Substituting the experimental data of Table~\ref{tab:values-metals}, we may compare with our numerical results.
Figures~\ref{fig:dragvarious-edge}, \ref{fig:dragvarious-screw} show good agreement of \eqref{eq:B-smallv}
below 40\%--50\% transverse sound speed (depending on the metal).
Finally, Table~\ref{tab:coefficients} lists the (dimensionless) values of the coefficients $f^{\txt{e,s}}_{0,1,2}/(8\m)^2$ for various metals.

\begin{table}[h!t!b]
{\renewcommand{\arraystretch}{1.1}
\centering
 \begin{tabular}{c|c|c|c|c}
          & Al\,(fcc) & Cu\,(fcc) & Fe\,(bcc) & Nb\,(bcc) \\\hline
 ${f_0^{\text{e}}}/{(8 \mu )^2}$ & 2.877 & 15.88 & 4.353 & 0.468 \\
 ${f_1^{\text{e}}}/{(8 \mu )^2}$ & 14.29 & 86.72 & 22.84 & 3.392 \\
 ${f_2^{\text{e}}}/{(8 \mu )^2}$ & 12.62 & 50.85 & 14.84 & 5.045 \\\hline
 ${f_0^{\text{s}}}/{(8 \mu )^2}$ & 2.723 & 22.15 & 6.172 & 0.747 \\
 ${f_1^{\text{s}}}/{(8 \mu )^2}$ & 23.72 & 192.1 & 53.63 & 6.562 \\
 ${f_2^{\text{s}}}/{(8 \mu )^2}$ & 55.94 & 464.6 & 128.1 & 14.98 \\
\end{tabular}
\caption{List of coefficients derived from second- and third-order elastic constants for various metals, as they appear in the small-velocity expansion of the drag coefficient, \eqnref{eq:B-smallv}.
All values in this list are dimensionless.}
\label{tab:coefficients}
}
\end{table}

\end{document}